\documentclass[preprint2]{aastex}
\usepackage{graphicx}
\usepackage{lscape}
\usepackage{units}
\usepackage{txfonts}
\usepackage{natbib}
\usepackage{color}
\usepackage{rotating} 
\def\hmpc{\ensuremath{h^{-1}}\,\textrm{Mpc}}
\def\hgpc{\ensuremath{h^{-1}}\,\textrm{Gpc}}

\def\degr{\hbox{$^\circ$}}
\def\magarcsec{\textrm{mag arcsec}\ensuremath{^{-2}}}

\def\abscostheta{\ensuremath{\left|\cos \theta\right|}}
\def\meanabscostheta{\ensuremath{\langle\left|\cos \theta\right|\rangle}}
\def\costheta{\ensuremath{\cos \theta}}

\def\sigmacostheta{\ensuremath{\sigma_{\abscostheta}}}
\def\sigmameanabscostheta{\ensuremath{\sigma_{\meanabscostheta}}}

\renewcommand{\vec}[1]{
\textbf{\textit{#1}}
}

\defcitealias{Lee2007a}{L07}
\defcitealias{Slosar2009}{S09}
\defcitealias{Trujillo2006}{T06}

\begin{document}
\title{The orientation of disk galaxies around large cosmic voids.}

\author{Jes\'us Varela}
\author{Juan Betancort-Rijo}
\author{Ignacio Trujillo}
\and
\author{Elena Ricciardelli}
\affil{
Instituto de Astrof\'{\i}sica de Canarias (IAC), E-38200 La Laguna, Tenerife, Spain\\
Depto. Astrof\'{\i}sica, Universidad de La Laguna (ULL), E-38206 La Laguna, Tenerife, Spain
}

\date{\today}

\begin{abstract}
  Using a large sample of galaxies from the SDSS-DR7, we have analysed
  the alignment of disk galaxies around cosmic voids. We have
  constructed a complete sample of cosmic voids (devoid of galaxies
  brighter than $M_r-5\log h=-20.17$) with radii larger than
  $10\,\hmpc$ up to redshift 0.12. Disk galaxies in shells around
  these voids have been used to look for particular alignments between
  the angular momentum of the galaxies and the radial direction of the
  voids. We find that disk galaxies around voids larger than $\gtrsim
  15\,\hmpc$ within distances not much larger than $5\,\hmpc$ from the
  surface of the voids present a significant tendency to have their
  angular momenta aligned with the void's radial direction
  with a significance $\gtrsim 98.8\%$ against the null
    hypothesis. The strenght of this alignment is dependent on the
  void's radius and for voids with $\lesssim 15\,\hmpc$ the
  distribution of the orientation of the galaxies is compatible with a
  random distribution. Finally, we find that this trend observed in
  the alignment of galaxies is similar to the one observed for the
  minor axis of dark matter halos around cosmic voids found in
  cosmological simulations, suggesting a possible link in the
  evolution of both components.
\end{abstract}

% 
%________________________________________________________________

%%%%%%%%%%%%%%%%%%%%%%%%%%%%%%%%%%%%%%%%%%%%%%%%%%%%%%%%%%%%%%%%%%%%%
%
% INTRODUCTION
%
%%%%%%%%%%%%%%%%%%%%%%%%%%%%%%%%%%%%%%%%%%%%%%%%%%%%%%%%%%%%%%%%%%%%%
\section{Introduction}

The study of the alignment of galaxies with respect to the large scale
structure is a recurrent topic still not fully settled.  The first
works studying the alignment of galaxies focused on clusters and
superclusters. Studies on the alignment of galaxies in
clusters~\citep{Adams1980} and
superclusters~\citep{Flin1986,Kashikawa1992} claimed to find
particular alignments of the galaxies with respect to their local
large scale structure. On the other side, similar studies did not find
any particular alignment~\citep{Helou1982, Dekel1985,
  Garrido1993}. More recently, \citet{Navarro2004} revisited the
analysis done by \citet{Flin1986} on the alignment of galaxies in the
Local Supercluster (LSC) under the light of the Tidal Torque
Theory~\citep[for a recent review about the Tidal Torque Theory or
TTT, see][]{Schaefer2009}. The authors found a tendency of galaxies to
have their spin parallel to the plane of the LSC, also known as
supergalactic plane, that would support the predictions from the TTT.

However, the observational analysis is hindered by two main
difficulties: the accurate determination of the direction of the
angular momentum of the galaxies and the determination of the
distribution of matter around them. The determination of the spin of
disk galaxies can be guessed by the shape of the galaxy, considering
that galaxies spin around their minor axis. However, there is still an
indetermination due to projection effects since in most of the cases
it is not possible to know which half, of the two in which a galaxy is
divided by its major axis, is closer to the observer. The presence of
dust lanes or the use of kinematic data can help to solve this
degeneracy but in most of the cases this information is not
available. To deal with this problem some authors have taken all the
possibilities of the spin as independent ones \citep{Kashikawa1992}
while others have opted for taking just one possibility
\citep{Lee2007a}.

Regarding the accurate determination of the mass distribution around
the galaxies, the main problem comes from the effects of the proper
motion of galaxies which introduces uncertainties in the conversion
from redshift to distances.

To overcome both problems, \citet[][hereafter T06]{Trujillo2006}
proposed the use of spiral galaxies seen edge-on or face-on (so the
direction of the spin vector is better determined) located in the
shells around cosmic voids. The advantage of the regions around large
cosmic voids is that the direction of the gradient of density is
strongly aligned with the radial direction which can be determined in
a robust way despite the uncertainties of converting redshifts in
distances. Using this technique and data from the third data release
of the Sloan Digital Sky Survey (SDSS-DR3) and the 2dF Galaxy Redshift
Survey (2dFGRS), \citetalias{Trujillo2006} found a  tendency of
galaxies around shells of voids to have their 
spin vector perpendicular to the radial
direction. \citet{Cuesta2008a} working on cosmological
  simulations of dark matter halos around voids found results in
  apparent agreement with those of \citetalias{Trujillo2006}. The
simulations show that the angular momentum of the dark matter halos
tend to be also aligned to the perpendicular direction. In both cases,
the results were in agreement with the prediction done using the TTT
\citep{Lee2000}, that the angular momentum would tend to be aligned
with the intermediate axis of the tidal shear tensor, that in the
surface of the voids is in the perpendicular direction.

However, recently, \citet[][hereafter S09]{Slosar2009} have redone a
similar analysis, but using a larger sample of galaxies from the
SDSS-DR6, obtaining a result that is consistent with a random
distribution of orientations, in contrast with the previous results.

In this work, we revisit the analysis of the alignment of galaxies
around voids with two significant improvements with respect to
those two previous works. First, we make use of the
latest data release of the SDSS, i.e. SDSS-DR7, and we combine it with
the morphological classification from the Galaxy Zoo
project~\citep{Banerji2010, Lintott2010} to select disk
galaxies. Second, we have developed a statistical procedure to
partially correct the indetermination in the spin direction due to the
projection effect so we can obtain information also from galaxies that
are not edge-on or face-on, increasing by a factor of 3 the effective
number of galaxies that are used in our analysis with respect to the
restriction to edge-on and face-on galaxies.

The outline of this Paper is as follows. Section~\ref{sec:data}
presents the data used for our analysis; Section~\ref{sec:cat_voids}
describes the procedure to search for voids; Section~\ref{sec:cat_gal}
is devoted to the selection of galaxies and the computation of their
alignments; Section~\ref{sec:results} contains the final results; in
Section~\ref{sec:discussion} the results are discussed and compared
with previous works and in Section~\ref{sec:summary} the summary of
the resuls are presented.

Through this paper we assume a $\Lambda$CDM cosmological model with
$\Omega_M=0.3$, $\Omega_\Lambda=0.7$ and \mbox{$H_0=100\,
  h$\,km\,s$^{-1}$\,Mpc$^{-1}$.}

%%%%%%%%%%%%%%%%%%%%%%%%%%%%%%%%%%%%%%%%%%%%%%%%%%%%%%%%%%%%%%%%%%%%%%
%
% DATA
%
%%%%%%%%%%%%%%%%%%%%%%%%%%%%%%%%%%%%%%%%%%%%%%%%%%%%%%%%%%%%%%%%%%%%%

\section{The data\label{sec:data}}

On what follows we describe the data that we have used to: a) create a
sample of cosmic voids, and b) obtain a sample of galaxies in the
shells surrounding them to explore the orientation of galaxies in the
large scale structure.

Our main source of data has been the New York University Value-Added
Galaxy Catalog%
\footnote{http://sdss.physics.nyu.edu/vagc/}%
\citep[NYU-VACG,][]{Blanton2005,Padmanabhan2008,Adelman-McCarthy2008},
which is based on the photometric and spectroscopic catalog of the
\mbox{SDSS-DR7}\footnote{http://cas.sdss.org/astrodr7/en/}~\citep{Strauss2002}. The
main characteristics of the NYU-VACG are:
\begin{itemize}
\item Spectroscopically complete up to $r\sim17.8$ (extinction
  corrected)
  \begin{itemize}
  \item Completeness $\sim 99\%$
  \item Success rate $\sim 99.9\%$
  \end{itemize}
\item $\mu_{50}(r-band)\le 24.5 \magarcsec$
\item $\sim 90$ targets/$\deg^2$
\item Median(z)=0.104
\end{itemize}

%%%%%%%%%%%%%%%%%%%%%%%%%%%%%%%%%%%%%%%%%%%%%%%%%%%%%%%%%%%%%%%%%%%%%%
% Selection of the subsample
%%%%%%%%%%%%%%%%%%%%%%%%%%%%%%%%%%%%%%%%%%%%%%%%%%%%%%%%%%%%%%%%%%%%%%
\subsection{Selection of the sample of galaxies}

We have established thresholds in absolute magnitude, $M_r^{lim}$, and
redshift, $z^{lim}$, in order to maximize the amount of galaxies while
keeping the final sample complete in spectroscopy. The completeness of
the initial catalog is a basic requirement to avoid the detection of
spurious voids.

\begin{figure}
  \centering
  \includegraphics[scale=0.38]{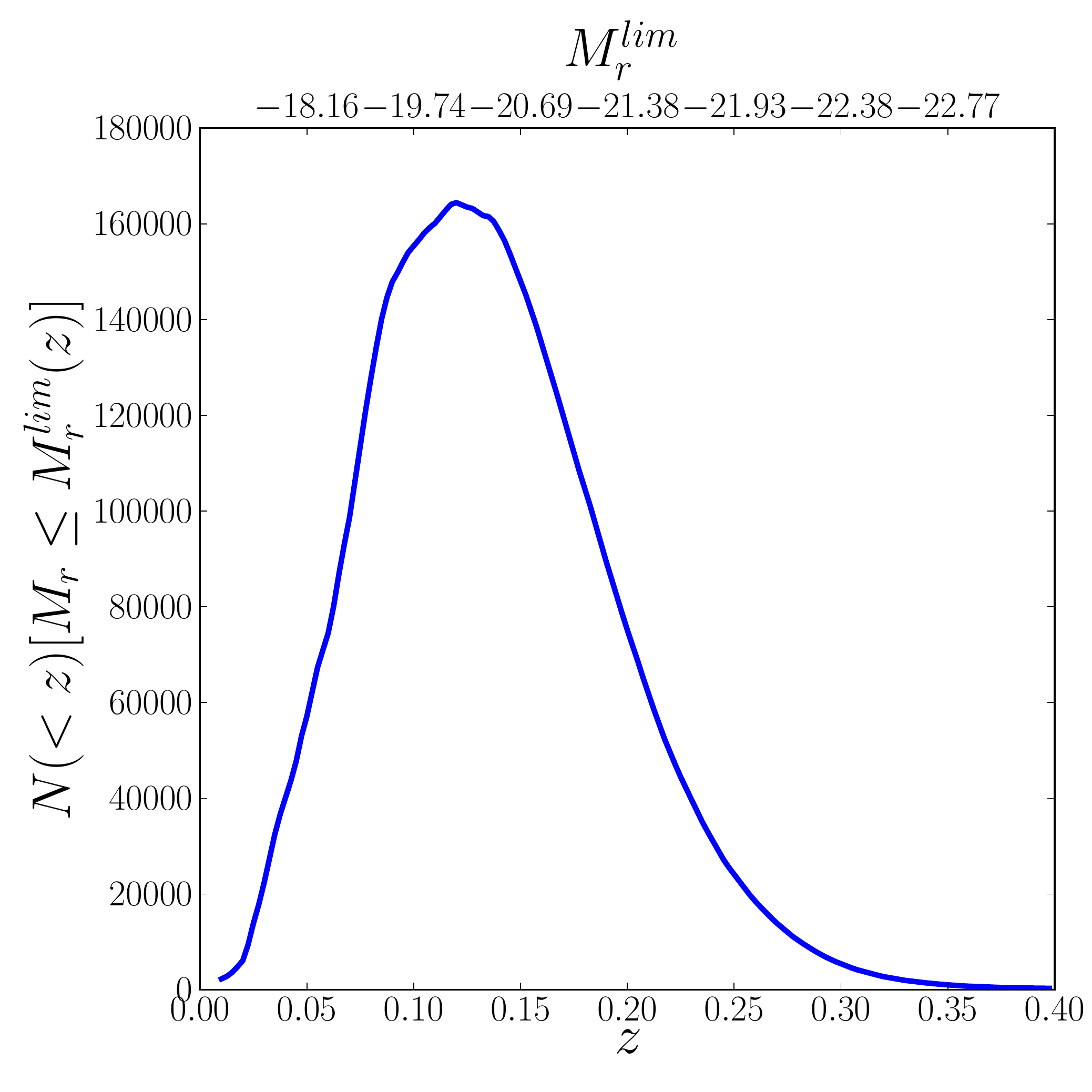}
  \caption{Number of galaxies brighter than $M_r^{lim}$ with
    $z<z^{lim}$. $M_r^{lim}$ is the absolute magnitude
    corresponding to the spectroscopic limit ($r=17.8$) at redshift
    $z^{lim}$. The peak of the distribution it is used to establish
    the thresholds in $M_r$ and $z$ of our initial
    catalog.\label{fig:plot1}}
\end{figure}

In Figure~\ref{fig:plot1} it is plotted the number of galaxies with
$M_r\leq M_r^{lim}(z)$ as a function of the redshift. The value of
$M_r^{lim}(z)$ corresponds to completeness limit ($r=17.8$) at each
redshift $z$. From the peak of this distribution we obtain the limits
of our final sample:
\begin{itemize}
\item $z\leq0.12$
\item $M_r-5\,\log h\leq-20.17$\footnote{We opted for no applying a
    k-correction due to the small redshift range probe and the high
    uncertainties in its determination.}
\end{itemize}

%%%%%%%%%%%%%%%%%%%%%%%%%%%%%
% Figure for M,z selection
%%%%%%%%%%%%%%%%%%%%%%%%%%%%%

%%%%%%%%%%%%%%%%%%%%%%%%%%%%%%%%%%%%%%%%%%%%%%%%%%%%%%%%%%%%%%%%%%%%%%
% Volume trimming
%%%%%%%%%%%%%%%%%%%%%%%%%%%%%%%%%%%%%%%%%%%%%%%%%%%%%%%%%%%%%%%%%%%%%%
\subsection{Volume trimming}

Although we focus our analysis in the largest continuous volume of
SDSS-DR7, the irregular limits of the volume still posed difficulties
in the reliable detection of voids. For this reason, we have defined
new regular limits minimizing the detection of spurious voids while
still keeping $\sim90\%$ of the original volume.

\begin{figure}[htb]
  \centering
  \includegraphics[scale=0.4]{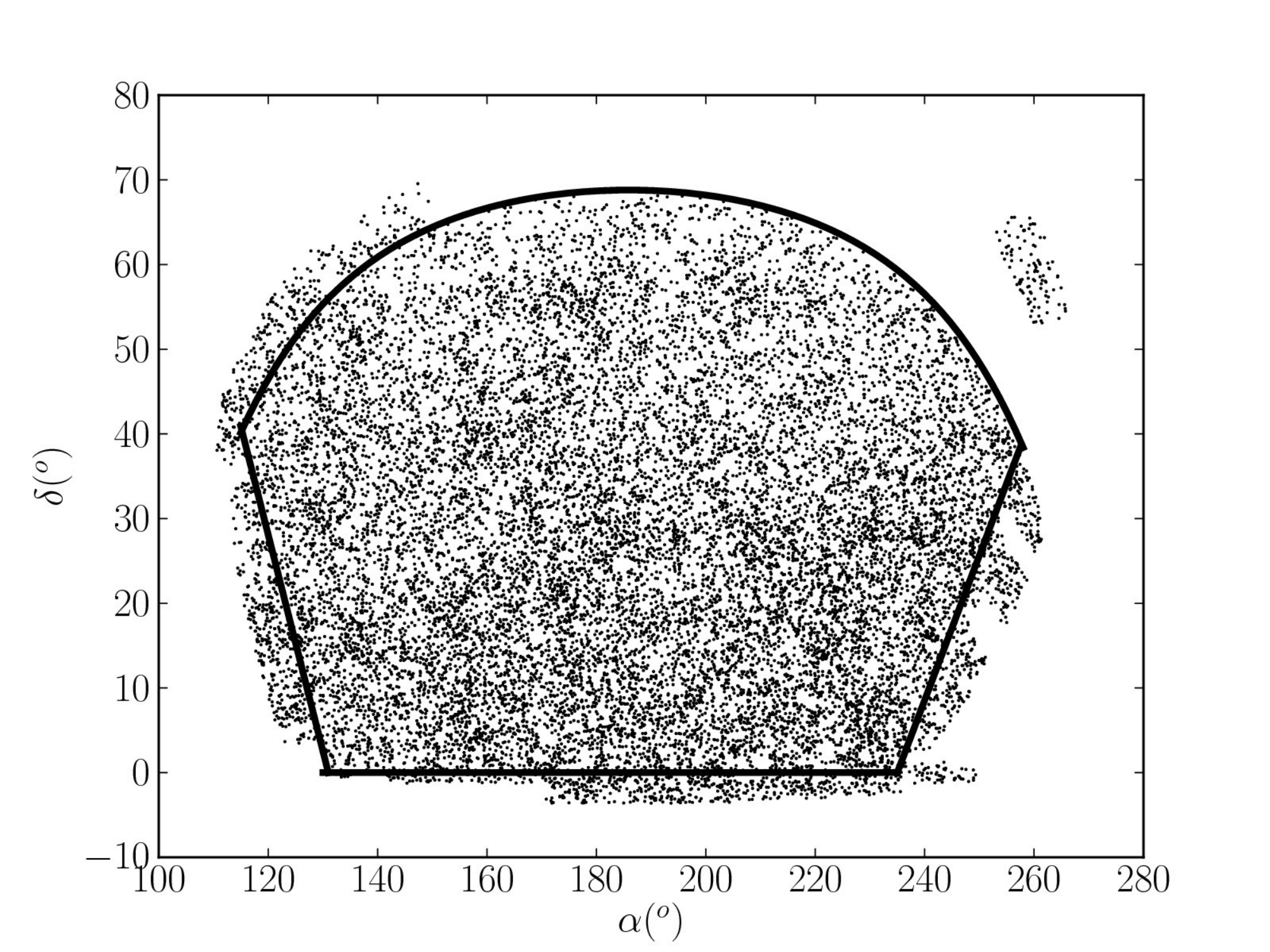}
  \caption{Partial projection of galaxies from the SDSS-NYU
    catalog. The limits used in this work are overplotted.\label{fig:nyuall}}
\end{figure}

Figure~\ref{fig:nyuall} shows the limits of our trimmed volume,
projected onto the original distribution of galaxies, defined as
follows:
\begin{itemize}
\item $\delta>0\degr$ [Southern limit]
\item $\delta< -2.555556\,(\alpha -
  131\degr)$ [Western limit]
\item $\delta< 1.70909\,(\alpha -
  235\degr)$ [Eastern limit]
\item $\delta<\arcsin\left[\frac{0.93232\,\sin(\alpha-95.9\degr)}
{\sqrt{1-[0.93232\,\cos(\alpha-95.9\degr)]^2}}\right]$ [Northern limit]
\end{itemize}

Table~\ref{tab:CatProp} summarizes the main properties of the
reference catalog that we have used in our search for cosmic voids.

\begin{table*}[btp]
  \centering
  \begin{tabular}[htp]{ll}
  \hline
  Reference catalog & NYU-VAGC (Galaxies) \\
  Spectroscopic completeness limit & $r\leq 17.8$ \\
  Redshift limits & $0.005 < z < 0.12$ \\
  Absolute magnitude limit & $M_r-5\,\log h\leq -20.17$\\
  & \\
  Number of galaxies & 142127 \\
  Total Projected Area & 1.941484 strad \\
  & $0.1545 \times 4\pi$ \\
  Total Volume & 0.0276556 (\hgpc)$^3$ \\
  Average density of galaxies & 0.00514 (\hmpc)$^{-3}$ \\
  & \\
  \hline
\end{tabular}
  \caption{Summary of properties of the reference catalog.}
  \label{tab:CatProp}
\end{table*}

%%%%%%%%%%%%%%%%%%%%%%%%%%%%%%%%%%%%%%%%%%%%%%%%%%%%%%%%%%%%%%%%%%%%%%
% Homogeinity Checks
%%%%%%%%%%%%%%%%%%%%%%%%%%%%%%%%%%%%%%%%%%%%%%%%%%%%%%%%%%%%%%%%%%%%%%
\subsection{Homogeinity check}

Another important requirement of the galaxy catalog to be suitable to
search for voids is the homogeinity. One common test of homogeinity is
the \mbox{$\langle V/V_{max}\rangle$} test, for which a value of 0.5
is expected for an homogenous distribution.

To perform this test, first of all, for each galaxy it is computed the
volume, $V$, of the sphere with radius the distance along the
line-of-sight to it. Then, the maximum of all the volumes, $V_{max}$,
is found and the ratio $V/V_{max}$ is obtained for each galaxy. The
final step is to calculate the average value of these ratios,
\mbox{$\langle V/V_{max}\rangle$}.
 
For our galaxy catalog \mbox{$\langle V/V_{max}\rangle=0.49990$},
confirming that at large scale the distribution of galaxies in our
volume is homogeneous.

%%%%%%%%%%%%%%%%%%%%%%%%%%%%%%%%%%%%%%%%%%%%%%%%%%%%%%%%%%%%%%%%%%%%%%
%
% DESCRIPTION OF THE PROCEDURE TO SEARCH FOR VOIDS
%
%%%%%%%%%%%%%%%%%%%%%%%%%%%%%%%%%%%%%%%%%%%%%%%%%%%%%%%%%%%%%%%%%%%%%

%%%%%%%%%%%%%%%%%%%%%%%%%%%%%%%%%%%%%%%%%%%%%%%%%%%%%%%%%%%%%%%%%%%%%%
% Procedure Description
%%%%%%%%%%%%%%%%%%%%%%%%%%%%%%%%%%%%%%%%%%%%%%%%%%%%%%%%%%%%%%%%%%%%%%
\section{Catalog of voids\label{sec:cat_voids}}

With the catalog of galaxies described in the previous section, we
proceed to search for cosmic voids on it. In this section we describe
the procedure followed to construct our catalog of voids.

\subsection{Procedure description}

First of all, we need to establish the definition of ``void'' that we
use in our analysis.  We have opted for the simplest one: a spherical
volume devoid of any galaxy brighter than our completeness limit. This
definition has been already used in other works such as
\citet{Patiri2006}, \citet{Patiri2006c}, \citet{Trujillo2006},
\citet{Brunino2007}, and \citet{Cuesta2008a}. \citet{Cuesta2008a}
found that for dark matter haloes in cosmological simulations, using
ellipsoidal voids instead of spherical ones does not affect
significantly their results. This gives us confidence in the use of
spherical voids for our analysis.

Apart from minor differences, the procedure that we have followed is
basically the \textsc{HB Void Finder} described in
\citet{Patiri2006}. These are the basic steps:
\begin{enumerate}
\item Random points are thrown within the volume of the catalog.
\item For each trial point, the 4 closest galaxies are found and the
  center and radius of the sphere defined by these 4 galaxies are
  computed and stored.
\item Of the resulting spheres, those fulfilling any of the following
  criteria are rejected:
  \begin{itemize}
  \item Not being empty.
  \item Intersecting the border of the
    volume.\footnote{\citet{Patiri2006} used a different
      technique. They put artificial galaxies in the limits of the
      survey, allowing voids to be defined by three real galaxies and
      one fake one.}
  \item Having a radius smaller than $10\,\hmpc$.
  \end{itemize}
\end{enumerate}

Finally, to have well defined spherical voids, we impose that they can
not overlap.  In the case in which several voids overlap, only the
largest one is kept. The order in which the rejection of the voids is
done affects their final sample. Therefore, to ensure that our final
sample contains the largest possible voids, the process of rejection
of overlapping voids it is done from the largest void to the smallest
one.

In summary, the voids selected for our final catalogs fulfill the
following conditions:
\begin{enumerate}
\item They are empty of galaxies from the initial catalog,
  i.e. $M_r<-20.17+5\,\log h$.
\item Their radius is larger than \unit[10]{\hmpc}.
\item They are completely inside the surveyed volume.%
  \footnote{Be aware that this criterion implies that the effective
    volume in which the center of voids can reside is smaller than the
    whole volume and depends on the sizes of the voids.}
\item Voids do not overlap, i.e. the distance between the centers of
    two voids is larger than the sum of their radii.
\end{enumerate}

The power of this procedure to produce a complete catalog of voids
depends critically on the relation between the density of galaxies,
the size of the voids and the number of trial points used in the
search. We have performed several tests and have found that using
\mbox{$\sim10^9$} initial random points (corresponding to a density of
\mbox{$\sim35\,(\hmpc)^{-3}$} trial points) ensures that the
completeness of our catalog of voids is $>99\%$.

The final outcome of this procedure is a complete catalog of 699
non-overlapping cosmic voids with radius larger than $10\,\hmpc$. We
found a median radius of $11.85\,\hmpc$ and the average density of
voids is
\mbox{$32.2\times10^{-3}\,(\hmpc)^{-3}$}.\footnote{Density
    computed within the effective volume in which the centers of the
    voids can be located.}

Table~\ref{tab:catvoids} provides the main information of these
voids. For each void, we include the position of the center, both in
Cartesian coordinates ($X,Y,Z$), and equatorial coordinates
($\alpha,\delta$) and redshift $z$; and the radius ($R$). The
Cartesian coordinates are computed as follows:
\begin{eqnarray*}
  \label{eq:CartesianCoords}
  X & = & D(z)\,\cos \delta\,\cos \alpha\\
  Y & = & D(z)\,\cos \delta\,\sin \alpha\\
  Z & = & D(z)\,\sin \delta
\end{eqnarray*}

where $D(z)$ is the comoving radial distance.

\begin{table*}
  \centering
  \begin{tabular}{r@{.}lr@{.}lr@{.}lr@{.}lr@{.}lr@{.}lr@{.}lr@{.}lr@{.}l}
\multicolumn{2}{c}{$X$} & 
\multicolumn{2}{c}{$Y$} & 
\multicolumn{2}{c}{$Z$} & 
\multicolumn{2}{c}{$R$} & 
\multicolumn{2}{c}{$\alpha$} & 
\multicolumn{2}{c}{$\delta$} & 
\multicolumn{2}{c}{$z$} \\
\multicolumn{2}{c}{[\hmpc]} & 
\multicolumn{2}{c}{[\hmpc]} & 
\multicolumn{2}{c}{[\hmpc]} & 
\multicolumn{2}{c}{[\hmpc]} & 
\multicolumn{2}{c}{[$\deg$]} & 
\multicolumn{2}{c}{[$\deg$]} &
\multicolumn{2}{c}{}\\
\hline\hline
-193&463 & -145&544 &  193&384 &   18&703 &  216&954492 &   38&617468 & 0&1059 \\
-227&081 &   26&372 &  140&238 &   18&205 &  173&375602 &   31&526741 & 0&0914 \\
-226&366 &   17&375 &  175&425 &   18&104 &  175&610699 &   37&692842 & 0&0979 \\
 -86&105 &   93&444 &  136&020 &   17&606 &  132&659407 &   46&949188 & 0&0630 \\
  \end{tabular}
  \caption{Catalog of voids (excerpt). Complete version in
    electronic form. See text for more details.}
  \label{tab:catvoids}
\end{table*}

%%%%%%%%%%%%%%%%%%%%%%%%%%%%%%%%%%%%%%%%%%%%%%%%%%%%%%%%%%%%%%%%%%%%%%
%
% Catalog of galaxies for measuring the alignment
%
%%%%%%%%%%%%%%%%%%%%%%%%%%%%%%%%%%%%%%%%%%%%%%%%%%%%%%%%%%%%%%%%%%%%%%

\section{Catalog of galaxies in shells around voids\label{sec:cat_gal}}

Using the previous catalog of voids, we extract those galaxies within
shells of 10\,\hmpc\ around each of them. The morphology of the
galaxies has been obtained from the Galaxy Zoo
catalog~\citep{Banerji2010, Lintott2010}, which provides robust
distinction between elliptical and disk galaxies, although $\sim52\%$
of the galaxies remain classified as ``uncertain''. We select only
those galaxies classified as ``spiral'' in the Galaxy Zoo catalog
($\sim37\%$) because for them the direction of the angular momentum is
well defined by their minor axes.

This results in a final sample of 45522 measurements, from 32374
single galaxies, that has been used to study the alignment of their
angular momentum with respect to the void.\footnote{These data is
  avalible upon request to the authors.} The use of measurements of
galaxies falling in more than one shell is justified because the small
uncertainties introduced are compensated by the increase in the size
of the sample. 

In the following sections we describe in detail how this alignment has
been computed.

%%%%%%%%%%%%%%%%%%%%%%%%%%%%%%%%%%%%%%%%%%%%%%%%%%%%%%%%%%%%%%%%%%%%%%
% Computation alignment
%%%%%%%%%%%%%%%%%%%%%%%%%%%%%%%%%%%%%%%%%%%%%%%%%%%%%%%%%%%%%%%%%%%%%%
\subsection{Computation of the alignment}

The advantage of studying galaxies around voids is that, on average,
the density increases radially. This makes the radial direction a good
proxy for the distribution of matter around each galaxy. Therefore, we
use the minor axis of the galaxies to define the orientation of their
angular momentum, and the radial direction of the voids to
characterize the distribution of matter around them. Hence, our
analysis is focused on the angle between these two directions,
$\theta$.\footnote{A schematic illustration of this method is shown in
  Figure 1 of \citetalias{Trujillo2006}.} For practical reasons, in
the context of this work we will use the expression ``radial
direction'' to mean the direction defined by a galaxy and the center
of the corresponding void, and ``perpendicular direction'' to mean any
direction perpendicular to the radial direction.

To compute the angle $\theta$ we need to define the direction of the
angular momentum or spin, $\vec s$, of each galaxy. We do this, first
of all, by computing the inclination angle between the plane of the
galaxy disk and the line of sight, $\zeta$. Following
\citet{Haynes1984} and \citet{Lee2007a}\footnote{\citet{Lee2007a} used
  the angle $i$ defined as the angle between the plane of the galaxy
  and the projected plane of the sky, therefore $\zeta=\pi/2-i$.}, we
use a model of thick disk with a projected minor-to-major axis ratio
$a/b$ and an intrinsic flatness $f$ (i.e. the ratio between the real
minor axis and the real major axis). According to this model, $\zeta$
can be obtained applying the formula:
\begin{equation}
  \label{eq:inclination}
  \sin^2 \zeta = \frac{(b/a)^2-f^2}{1-f^2}
\end{equation}

For values of $b/a<f$, the angle is set to 0.

The flatness $f$ depends on the morphological type of the galaxies and
we use an average value of 0.14.

From Equation~(\ref{eq:inclination}) it is easy to see that to a
single value of $b/a$ corresponds two values of $\zeta$:
$\zeta_+=|\zeta|$ and $\zeta_-=-|\zeta|$. The indetermination is
irrelevant for $\zeta=0$ (edge-on galaxies) and for $\zeta=\pm\pi/2$
(face-on galaxies).

To compute the spin vector $\vec s$ we followed the prescription
by~\citetalias{Trujillo2006}. According to this, if ($\alpha$,$\delta$) are
the equatorial coordinates of a galaxy, $\zeta$ the inclination angle
obtained from Equation~(\ref{eq:inclination}) and $\phi$ is the position angle
of the galaxy increasing counterclockwise (i.e. from north to east in the
plane of the sky), the components of $\vec s$ are:

\begin{eqnarray}
  \label{eq:spin_comp}
  s_x & = & \cos \alpha \cos \delta \sin \zeta \nonumber\\
      &   & + \cos \zeta (\sin \phi \cos \alpha
  \sin \delta - \cos \phi \sin \alpha)\\*
s_y & = & \sin \alpha \cos \delta \sin \zeta \nonumber\\
 &&+ \cos \zeta (\sin \phi
\sin \alpha \sin \delta + \cos \phi \cos \alpha)\\
s_z & = & \sin \delta \sin \zeta - \cos \zeta \sin \phi \cos \delta
\end{eqnarray}

Next, we compute the angle between the radial vector that connects the
center of the void, $\vec r_{void}$, with the center of the galaxy, $\vec
r_{galaxy}$:
\begin{equation}
  \vec r = \vec r_{galaxy}- \vec r_{void}.
\end{equation}

Having obtained $\vec r$ and $\vec s$, the angle between them,
$\theta$, it is computed as:

\begin{equation}
  \label{eq:cos_theta}
  \theta = \arccos \left( \frac{\vec s \cdot \vec r}{|\vec s||\vec r|} \right)
\end{equation}

%%%%%%%%%%%%%%%%%%%%%%%%%%%%%%%%%%%%%%%%%%%%%%%%%%%%%%%%%%%%%%%%%%%%%%
% Characterization of the cos(theta)
%%%%%%%%%%%%%%%%%%%%%%%%%%%%%%%%%%%%%%%%%%%%%%%%%%%%%%%%%%%%%%%%%%%%%%
\subsection{Analytical model of the distribution of $\theta$}

We compute the angle $\theta$ for all the galaxies in our sample of
galaxies around voids, obtaining a distribution of
$\theta$. \citet{Betancort-Rijo2009} provide an analytical model for
the distribution of the angle $\theta$, or to be more precise, of
\abscostheta, $P(\abscostheta)$. From theoretical principles confirmed
by simulations, the authors found that $P(\abscostheta)$ is well
described by the expression:
\begin{equation}
  \label{eq:pcostheta}
  P(\mu) = \frac{p\,du}{[1+(p^2-1)\mu^2]^{3/2}};\, \mu\equiv \abscostheta,
\end{equation}
where $p$ is a free parameter that describes the overall shape of the
probability distribution. An interesting property of this distribution
is the relation between the parameter $p$ and the average value of
$\abscostheta$:

\begin{equation}
  \label{eq:p_from_cos}
  p = \frac{1}{\meanabscostheta} - 1
\end{equation}

Besides, the values of $p$ are related with the existence or absence
of particular alignment between the vectors $\vec r$ and $\vec s$\
according to the following criteria:
\begin{description}
\item[$p<1$.] $\vec r$ and $\vec s$\ tend to be parallel.
\item[$p=1$.] There is no particular alignment between $\vec r$ and
  $\vec s$.
\item[$p>1$.] $\vec r$ and $\vec s$\ tend to be perpendicular.
\end{description}

It has been found that Equation (\ref{eq:pcostheta}) describes well
the results from cosmological simulations \citep{Brunino2007,
  Cuesta2008a}.

An alternative expression for the same distribution described by
Equation (\ref{eq:pcostheta}) it is provided by~\citet{Lee2004} and
used in several works~(eg., \citetalias{Trujillo2006},
\citeauthor{Lee2007a} 2007,
  \citetalias{Slosar2009}). This alternative expression is characterized by a
parameter $c$ and, from the comparison between Equation (9) of
\citetalias{Slosar2009} and Equation~(\ref{eq:pcostheta}) of the present
work, it is possible to obtain the following expression relating both
characteristic parameters:
\begin{equation}
  \label{eq:c_from_p}
  p=\sqrt{1+\frac{3c}{2(1-c)}}
\end{equation}

%%%%%%%%%%%%%%%%%%%%%%%%%%%%%%%%%%%%%%%%%%%%%%%%%%%%%%%%%%%%%%%%%%%%%%
% Statistical correction
%%%%%%%%%%%%%%%%%%%%%%%%%%%%%%%%%%%%%%%%%%%%%%%%%%%%%%%%%%%%%%%%%%%%%%
\subsection{Statistical computation of $P(\abscostheta)$}

Different approaches have been used to deal with the
  indetermination of the values of $\zeta$. For example,
  \citet{Kashikawa1992} uses the two values of $\zeta$
  independently. Another possibility is to use just one sign in the
  definition of $\zeta$ as done by \citet{Lee2007a}. However, these
  authors acknowledge that this decreases the strength of the measured
  alignment.

On the other hand, \citetalias{Trujillo2006} and \citetalias{Slosar2009} have overcome
the problem with the indetermination of the values of $\zeta$ using
only edge-on and face-on galaxies, for which the direction of the spin
is well determined. The main disadvantage of this approach is that the
number of galaxies suitable for computing $P(\abscostheta)$ is greatly
reduced. For example, using the criteria
from~\citetalias{Trujillo2006}, the fraction of galaxies that can be
used is $\sim 22\%$ of all the disk galaxies.

We opted for a statistical approach that allows to compute a corrected
distribution $P_c(\abscostheta)$ from the combination of the
distributions $P(\abscostheta)$ obtained using both signs.

In Appendix~\ref{sec:appendixA} we describe in detail this
procedure. We also show in this appendix the results from several
Monte Carlo simulations that show the ability of the procedure to
recover the correct values of $p$ (Table~\ref{tab:p_check}).

The fact that we actually do not know the real values of $\zeta$ is
reflected in the uncertainties of the procedure. Although the values
of $p$ are well recovered, the uncertainties measured from the
simulations are larger than those expected from considering just the
size of the sample, $N_g$. Of course, this is because we are not using
the real values of $\zeta$. Nevertheless, from the simulations
we have obtained that our procedure has a predictibility power
equivalent to that of a sample $0.6\,N_g$ with complete knowledge of
the real values of $\zeta$. Let's remember that the common procedure
of using only face-on or edge-on galaxies is restricted to $\sim 20\%$
of the total amount of spiral galaxies. This means that our
statistical procedure increases by a factor of 3 the effective number
of galaxies with respect to previous works.

%%%%%%%%%%%%%%%%%%%%%%%%%%%%%%%%%%%%%%%%%%%%%%%%%%%%%%%%%%%%%%%%%%%%%%
% 
% Results
%
%%%%%%%%%%%%%%%%%%%%%%%%%%%%%%%%%%%%%%%%%%%%%%%%%%%%%%%%%%%%%%%%%%%%%%

\section{Results\label{sec:results}}

Using the procedure described in Appendix~\ref{sec:appendixA}, we have
computed the corrected distribution $P_c(\abscostheta)$ of the sample
of galaxies around voids. Given the large size of our initial sample,
we have also computed $P_c(\abscostheta)$ for different subsamples
combining different sizes of voids and shells around them.

An important point of our analysis has been to establish the
significance of our results in a robust way. This is done by
comparing, in each case, the measured signal with the standard
deviation \sigmameanabscostheta\ of the theoretical distribution in case of null
signal, i.e. $\meanabscostheta=0.5$.\footnote{In the analysis of the
  significance is more convenience the use of \meanabscostheta\ than
  that of $p$ because the former has a gaussian distribution but the
  latter has not.} In the situation of complete knowledge of the real
values of $\zeta$, we would have $\sigmameanabscostheta=(\sqrt{12\times N_g})^{-1}$ for
a sample of $N_g$ measures.  However, we have already shown that our
procedure has uncertainties equivalent to a sample of size $0.6\,N_g$,
therefore, the previous expression needs to be corrected to

\begin{equation}
  \label{eq:sigmatheta}
  \sigmameanabscostheta=\frac{1}{\sqrt{12 \times 0.6 \times N_g}}.  
\end{equation}

Knowing the value of \sigmameanabscostheta, we can establish the signal to noise
ratio ($SNR$) of the signal of a subsample of $N_g$ measurements as:

\begin{equation}
  \label{eq:snr}
  SNR = \frac{0.5-\meanabscostheta_{corr}}{\sigmameanabscostheta},
\end{equation}

where $\meanabscostheta_{corr}$ is obtained from the statistical
correction and \sigmameanabscostheta\ from
Equation~(\ref{eq:sigmatheta}). Note that the denominator is the
signal, which corresponds to the difference between the observed value
of $\meanabscostheta_{corr}$ and that of the random distribution which
is 0.5. For practical reasons, the sign of $SNR$ has been chosen so
that is positive for values of $p>1$ ($\meanabscostheta_{corr}<0.5$)
and negative for values of $p<1$ ($\meanabscostheta_{corr}>0.5$).

In Tables~\ref{tab:p_rvoid_shellwidth} and
\ref{tab:p_rvoid_shellwidth_dif} are shown the main results of our
analysis. In Table~\ref{tab:p_rvoid_shellwidth}, samples are
constructed by setting a minimum value for the radii of the voids
($R_{Void}^{min}$) while, in Table~\ref{tab:p_rvoid_shellwidth_dif},
samples are constructed using voids with radii in the ranges
$R_{Void}\pm 0.5\,\hmpc$.\footnote{For convenience, we will refer the
  samples of the first table as ``cumulative'' samples and those of
  the second table as ``differential'' samples.} Apart from this
difference in the definition of the first column, both tables share
the description of the rest of the columns: $SW$ is the width of the
innermost shell in $\hmpc$; $N$ is the number of measures of the
sample; \meanabscostheta\ is the mean of the \abscostheta; $p$ is the
characteristic parameter of Equation~(\ref{eq:pcostheta}); and $SNR$
is the signal to noise ratio computed with
Equation~(\ref{eq:snr}). From simulations it has been
  found that results obtained with less than $\sim100$ measures are
  not reliable, therefore, samples with less than this number have
  been flagged with a question mark beside the value of the $SNR$.

Errors in $\meanabscostheta$ are computed using the standard deviation
of $\abscostheta$ resulting from 10000 Monte Carlo simulations with no
signal (see below) corrected using Equations~(\ref{eq:err_cos_theta_1}-\ref{eq:err_cos_theta_3})
assuming the following relation:

\begin{equation}
  \label{eq:errcos_sim}
\frac{\sigma_{\meanabscostheta} (p \neq 1;corr)}{\sigma_{\meanabscostheta} (p\neq1;theo)} =
\frac{\sigma_{\meanabscostheta} (p=1;sim)}{\sigma_{\meanabscostheta} (p=1;theo)}
\end{equation}

where $\sigma_{\meanabscostheta}(p;theo)$ is computed using the
theoretical expressions described in
Equations~(\ref{eq:err_cos_theta_1}-\ref{eq:err_cos_theta_3}), 
$\sigma_{\meanabscostheta}(p=1;sim)$ is computed from the simulations
and $\sigma_{\meanabscostheta} (p \neq 1;corr)$ is the final value
used in Tables~\ref{tab:p_rvoid_shellwidth}-\ref{tab:p_rvoid_shellwidth_dif}.

Since $p$ does not follow a Gaussian distribution, for this parameter
we provide the confidence levels at $1\sigma$ again correcting the
theoretical values from
Equations~(\ref{eq:p_errors_1}-\ref{eq:p_errors_2}) with the
confidence levels measured from the simulations with no signal.

\begin{figure*}
  \centering
  \includegraphics[scale=0.55]{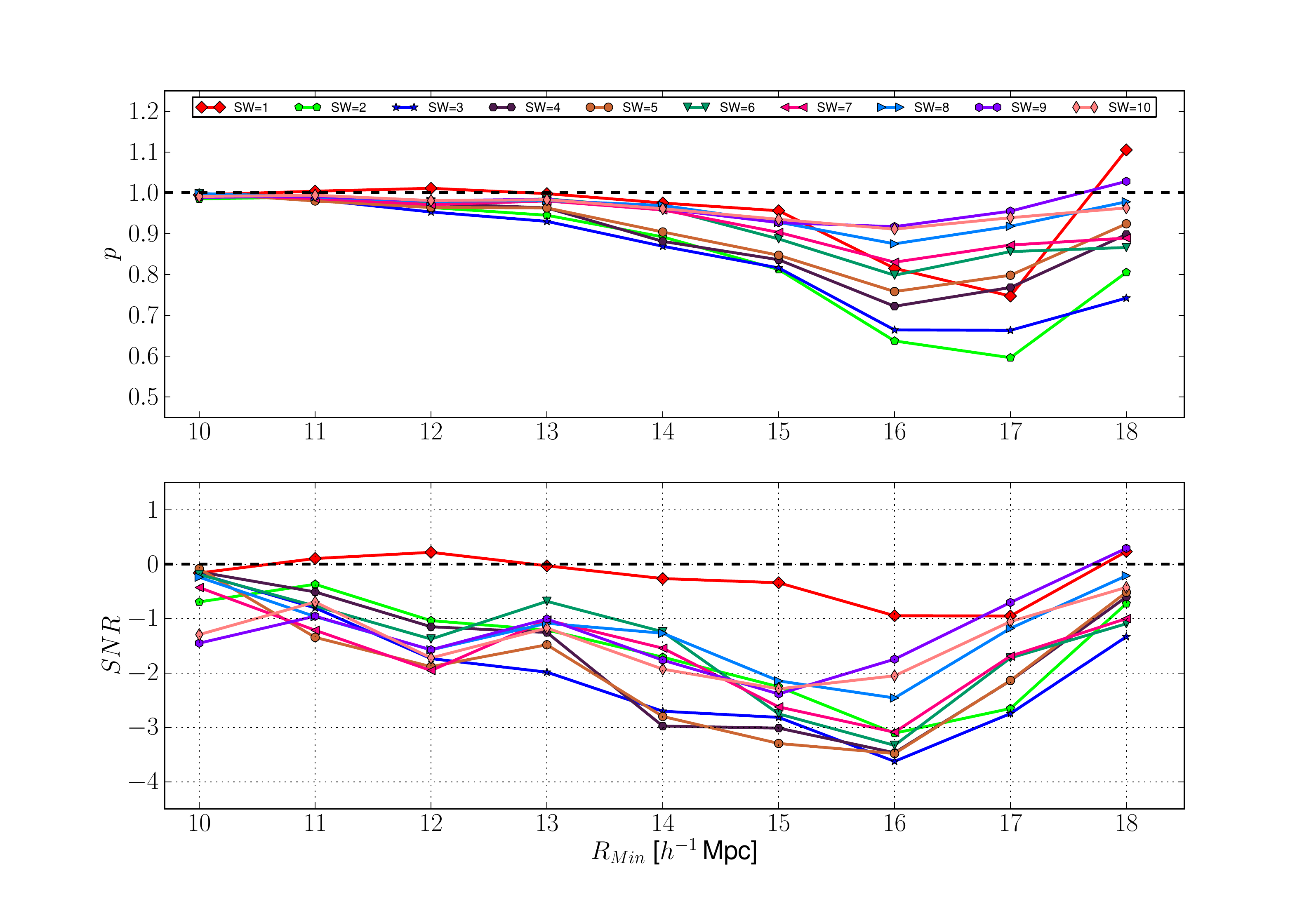}
  \caption{\textit{Upper panel:} Values of the parameter $p$ for
    subsamples of galaxies in shells of width $SW$ and voids with
    radius larger than $R_{Min}$. \textit{Lower panel:} Signal to
    noise of the alignment found for each subsample. This gives the
    significance of rejecting the null hyphotesis of not existence of
    any alignment.\label{fig:signal}}
\end{figure*}

\begin{figure*}
  \centering
  \includegraphics[scale=0.55]{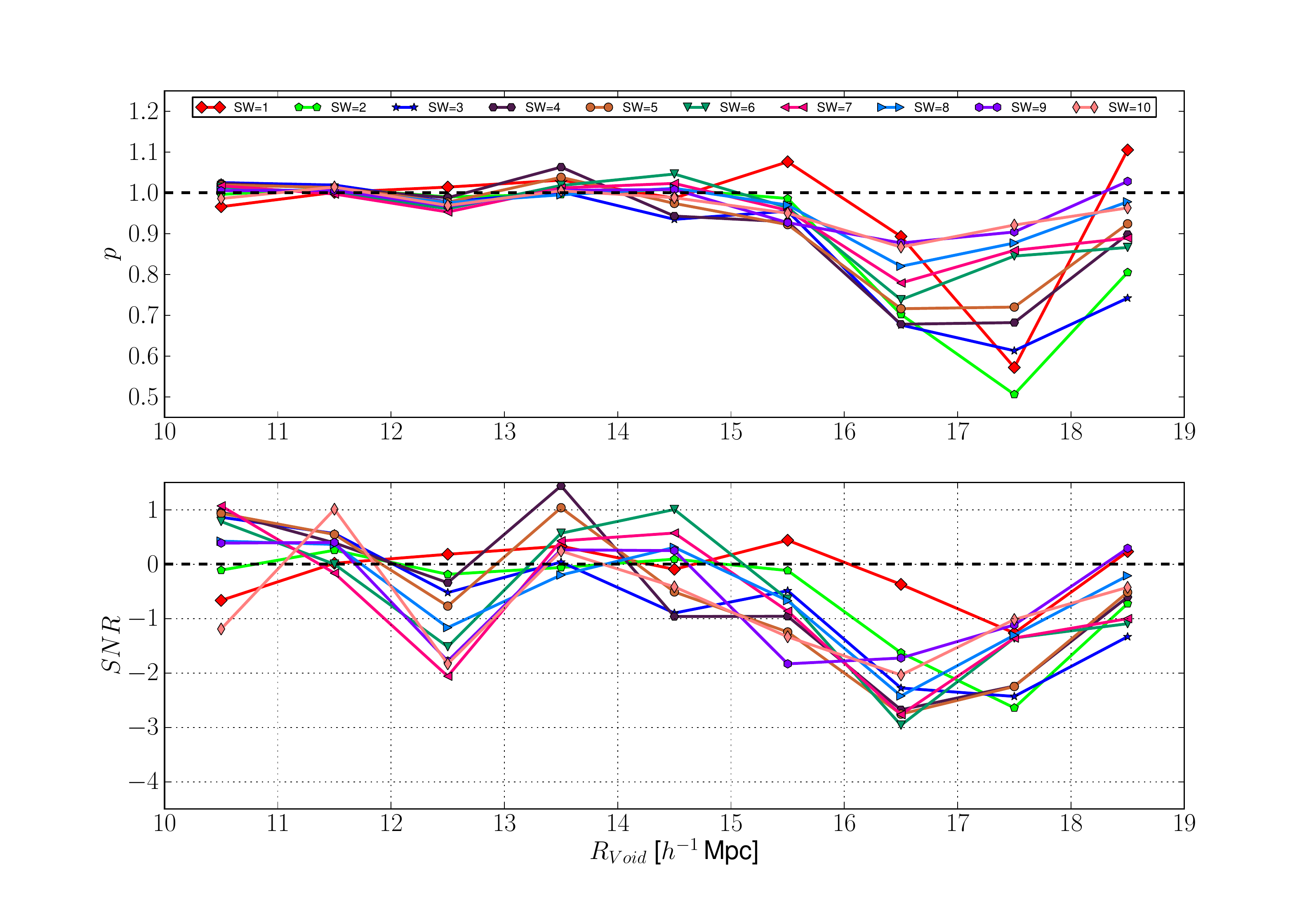}
  \caption{\textit{Upper panel:} Values of the parameter $p$ for
    subsamples of galaxies in shells of different widths $SW$ (in
    \hmpc) and voids in bins of $1\,\hmpc$ in radius. \textit{Lower
      panel:} Signal to noise of the alignment found for each
    subsample. This gives the significance of rejecting the null
    hyphotesis of not existence of any
    alignment.\label{fig:signal_differential}}
\end{figure*}

In Figures~\ref{fig:signal} and \ref{fig:signal_differential} are
plotted the values of $p$ (upper panels) and $SNR$ (lower panels) as a
function of the radius of the voids, for the cumulative and
differential samples, respectively. In the first figure the plotted
radius is the minimum radius of each sample and in the second the
central radius of each bin. In different colors are plotted the 10
shells widths that have been explored.

The first result is that $p<1$ for most of the subsamples. This means
that the direction of the spin of the galaxies tends to be parallel to
the radial direction of the void. Despite of the large sample that we
are using and the additional statistical correction, we find that the
significance of the signal is not high most of the times and is
dependent on the radius of the voids. The highest significance is
reached when selecting galaxies around voids, larger than $16\,\hmpc$
in shells of $3\,\hmpc$ ($|SNR|>3.6$) using 179 galaxies, but the
$|SNR|$ is higher than 3 increasing the width of the shell up to
$7\,\hmpc$ and the sample size to 614 galaxies. Therefore, the further
the galaxies are from the surface of the void, the lower is the signal
although the increase in the size sample keeps the significance high.

The relation between the strength of the alignment and the radius of
the void is clearly shown in Figure~\ref{fig:signal_differential}
where the differential samples are used. For voids smaller than
$\sim15\,\hmpc$ these results are compatible with a random
distribution. It is for voids $\gtrsim15\,\hmpc$ when it appears a
signal of alignment, although given the smaller size of the samples
the highest significance reached is 2.96 for voids with
$16\,\hmpc\leq R_{Void}\leq 17\,\hmpc$ and a shell of $6\,\hmpc$.

Figure~\ref{fig:Pcorr_hist} shows the corrected histograms of $\theta$
values (see Equation~\ref{eq:p_corr}) for the cases in which is
reached the maximum $SNR$ for the cumulative (left panel) and the
differential samples (right panel). The continuous red line shows the
analytical model described by Equation~(\ref{eq:pcostheta}) with the
$p$ values corresponding to these two maxima.

To compute in a more robust way the significance level of
  our results we need to compare them with a control sample with no
  signal. To construct this control sample, we have run 10000 Monte
  Carlo simulations in which the spin direction of the galaxies
  (determined by their position angles and axial ratios) has been
  shuffled so that each galaxy it is assigned the spin direction of any
  other galaxy randomly selected. This procedure has the advantage of
  ensuring the randomness of the spin distribution and, therefore, the
  lack of any alignment signal, while using real data.

For each simulation we have repeated the analysis
  performed in the real data using cumulative subsamples, and we have
  also computed 2 statistics used both in the real data and in the
  simulations. These statistics were computed as follows: for each bin
  in $R_{Void}$, the mean (median) value of the $SNR$ measured in the
  10 different shell widths was computed and the extreme value
  (i.e. with highest absolute value) of each simulation was kept.

The significance level from each statistic is computed
  as

  $1-2f(\textrm{extreme}\{SNR_{Sim}\}<\textrm{extreme}\{SNR_{Real}\})$,

  where
    $f(\textrm{extreme}(SNR_{Sim})<\textrm{extreme}\{SNR_{Real}\})$ is
    the fraction of simulations with extreme values of the mean
    (median) of SNR lower than the observed ones (-2.73 for the mean;
    -3.10 for the median). The fact of multiplying by two the observed
    fraction takes into account that we are considering only one side
    of the distribution of the statistics.

The results from this analysis are summarized in
  Table~\ref{tab:final_sim}. Using the mean we obtain a significance
  of $98.8\%$ which improves slightly to $99.5\%$ if the median is
  used instead.

\begin{figure*}
  \centering
  \includegraphics[scale=0.55]{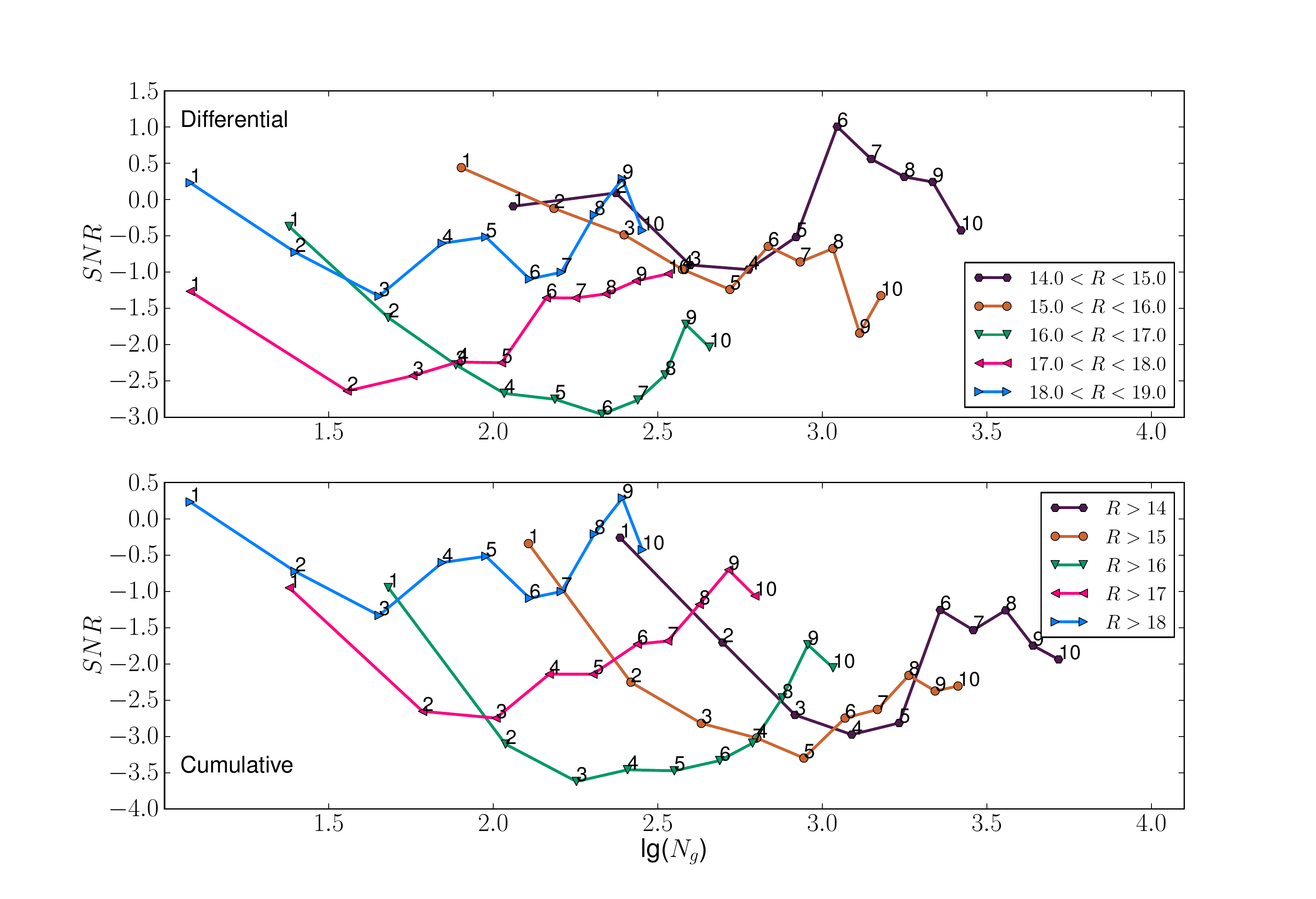}
  \caption{Variation of the SNR as a function of the number of
    galaxies $N_g$ for different samples. Each line corresponds to a
    selection in void's radius (differential in the upper panel and
    cumulative in the lower one) and each number indicates the width
    of the shell in $\hmpc$. For practical purposes, only the samples
    of voids larger than $14\,\hmpc$ are
    shown.\label{fig:snr_vs_ngal}}
\end{figure*}

Although this test ensures the existence of a global signal, we have
also checked that the increase of the $SNR$ with the radius of the
voids shown in Tables~\ref{tab:p_rvoid_shellwidth} and
\ref{tab:p_rvoid_shellwidth_dif} is not just a consequence of the
variation in the size of the samples.  Figure~\ref{fig:snr_vs_ngal}
shows the dependence of the $SNR$ with the size of the samples
($N_g$). For clarity purposes we have restricted the analysis to the
voids larger than $14\,\hmpc$. Each line corresponds to different
limits in $R_{Void}$ and each point corresponds to a shell width. It
can be seen that all subsamples show a similar trend with the width of
the shell, showing the maximum value of $SNR$, in absolute value, at
intermediate shell widths. Also, for subsamples with similar sizes,
the $SNR$ shows variations depending on the size of the voids and
shells, rejecting the hypothesis that the variations of the $SNR$ are
due only to variations of the size of the samples.

\begin{table}
  \centering
  \begin{tabular}{lcc}
    Criteria & $N_{Sim}$ & \% \\
\hline\hline
$\min\{\langle SNR\rangle\}<-2.73$ & 58 & 98.8 \\
$\min\{\textrm{median}(SNR)\}<-3.10$ & 23 & 99.5\\
  \end{tabular}
  \caption{Results from 10000 simulations with reshuffling of position
    angle and axial ratio between galaxies, showing the number 
    of simulations presenting
    values for two statistics smaller than the observed
    values ($\textrm{extreme}\{\langle SNR\rangle\}$(Real Data)=-2.73;
    $\textrm{extreme}\{\textrm{median}(SNR)\}$(Real
    data)=-3.10). The last column shows the significance of the
    result, which takes into account that we are considering only
    one side of the distribution of the statistics. See text for more details.}
  \label{tab:final_sim}
\end{table}

Finally, to check the dependence of the significance of the signal
with the distance of the galaxies to the surface of the void, we have
constructed subsamples containing galaxies closer than $5\,\hmpc$ of
the voids' surface and galaxies between $5\,\hmpc$ and
$10\,\hmpc$. With these two groups of galaxies we have repeated the
analysis (only cumulative). In Table~\ref{tab:signal_two_shells} are
presented the results of this analysis which show that galaxies at
distances larger than $5\,\hmpc$ do not present any significance
alignment.

%%%%%%%%%%%%%%%%%%%%%%%%%%%%%%%%%%%%%%%%%%%%%%%%%%
% TABLE: Results, no overlapping, cumulative
%%%%%%%%%%%%%%%%%%%%%%%%%%%%%%%%%%%%%%%%%%%%%%%%%%

\begin{sidewaystable*}
\scriptsize
\footnotesize
  \centering
\begin{tabular}{ll}

  \begin{tabular}{cccccc}
$R_{Void}^{min}$ & $SW$ & $N$ & $\meanabscostheta$ & $p$ & SNR
\\
\hline\hline
 10  &  1  & 2320 & 0.501 $\pm$ 0.009 & $0.995_{-0.033}^{+0.032}$ & -0.157 \\
 11  &  1  & 1497 & 0.499 $\pm$ 0.011 & $1.004_{-0.043}^{+0.043}$ & 0.106 \\
 12  &  1  & 884 & 0.497 $\pm$ 0.014 & $1.011_{-0.056}^{+0.054}$ & 0.211 \\
 13  &  1  & 503 & 0.500 $\pm$ 0.019 & $0.998_{-0.072}^{+0.072}$ & -0.026 \\
 14  &  1  & 243 & 0.506 $\pm$ 0.026 & $0.975_{-0.101}^{+0.103}$ & -0.260 \\
 15  &  1  & 128 & 0.511 $\pm$ 0.034 & $0.956_{-0.134}^{+0.134}$ & -0.341 \\
 16  &  1  &  48 & 0.551 $\pm$ 0.057 & $0.815_{-0.197}^{+0.180}$ & -0.948 ?\\
 17  &  1  &  24 & 0.572 $\pm$ 0.085 & $0.747_{-0.311}^{+0.185}$ & -0.951 ?\\
 18  &  1  &  12 & 0.475 $\pm$ 0.145 & $1.105_{-1.172}^{+0.029}$ & 0.232 ?\\
\hline
 10  &  2  & 4683 & 0.504 $\pm$ 0.005 & $0.985_{-0.023}^{+0.024}$ & -0.687 \\
 11  &  2  & 2998 & 0.502 $\pm$ 0.008 & $0.990_{-0.030}^{+0.031}$ & -0.355 \\
 12  &  2  & 1781 & 0.509 $\pm$ 0.010 & $0.964_{-0.037}^{+0.037}$ & -1.032 \\
 13  &  2  & 1015 & 0.514 $\pm$ 0.013 & $0.945_{-0.050}^{+0.049}$ & -1.220 \\
 14  &  2  & 498 & 0.528 $\pm$ 0.018 & $0.892_{-0.065}^{+0.067}$ & -1.705 \\
 15  &  2  & 262 & 0.552 $\pm$ 0.024 & $0.812_{-0.082}^{+0.081}$ & -2.252 \\
 16  &  2  & 109 & 0.611 $\pm$ 0.036 & $0.637_{-0.101}^{+0.096}$ & -3.106 \\
 17  &  2  &  61 & 0.627 $\pm$ 0.050 & $0.596_{-0.132}^{+0.121}$ & -2.653 ?\\
 18  &  2  &  25 & 0.554 $\pm$ 0.117 & $0.805_{-0.616}^{+0.048}$ & -0.725 ?\\
\hline
 10  &  3  & 7592 & 0.501 $\pm$ 0.004 & $0.997_{-0.019}^{+0.018}$ & -0.185 \\
 11  &  3  & 4877 & 0.504 $\pm$ 0.006 & $0.983_{-0.024}^{+0.023}$ & -0.827 \\
 12  &  3  & 2888 & 0.512 $\pm$ 0.008 & $0.953_{-0.030}^{+0.030}$ & -1.751 \\
 13  &  3  & 1663 & 0.518 $\pm$ 0.010 & $0.930_{-0.038}^{+0.038}$ & -1.979 \\
 14  &  3  & 826 & 0.535 $\pm$ 0.014 & $0.869_{-0.050}^{+0.049}$ & -2.703 \\
 15  &  3  & 429 & 0.551 $\pm$ 0.019 & $0.816_{-0.064}^{+0.063}$ & -2.819 \\
 16  &  3  & 179 & 0.601 $\pm$ 0.029 & $0.664_{-0.081}^{+0.081}$ & -3.620 \\
 17  &  3  & 102 & 0.601 $\pm$ 0.040 & $0.663_{-0.115}^{+0.112}$ & -2.747 \\
 18  &  3  &  45 & 0.574 $\pm$ 0.087 & $0.742_{-0.362}^{+0.112}$ & -1.331 ?\\
\hline
 10  &  4  & 11060 & 0.501 $\pm$ 0.005 & $0.998_{-0.016}^{+0.015}$ & -0.148 \\
 11  &  4  & 7083 & 0.502 $\pm$ 0.005 & $0.991_{-0.021}^{+0.019}$ & -0.521 \\
 12  &  4  & 4232 & 0.507 $\pm$ 0.007 & $0.974_{-0.025}^{+0.025}$ & -1.137 \\
 13  &  4  & 2463 & 0.509 $\pm$ 0.009 & $0.963_{-0.033}^{+0.031}$ & -1.265 \\
 14  &  4  & 1229 & 0.532 $\pm$ 0.012 & $0.881_{-0.041}^{+0.042}$ & -2.974 \\
 15  &  4  & 632 & 0.545 $\pm$ 0.016 & $0.836_{-0.054}^{+0.054}$ & -3.021 \\
 16  &  4  & 256 & 0.581 $\pm$ 0.025 & $0.722_{-0.073}^{+0.075}$ & -3.459 \\
 17  &  4  & 148 & 0.566 $\pm$ 0.034 & $0.768_{-0.104}^{+0.103}$ & -2.141 \\
 18  &  4  &  70 & 0.527 $\pm$ 0.060 & $0.898_{-0.233}^{+0.183}$ & -0.603 ?\\
\hline
 10  &  5  & 15289 & 0.500 $\pm$ 0.003 & $0.999_{-0.014}^{+0.013}$ & -0.053 \\
 11  &  5  & 9845 & 0.505 $\pm$ 0.004 & $0.980_{-0.017}^{+0.017}$ & -1.365 \\
 12  &  5  & 5865 & 0.509 $\pm$ 0.006 & $0.964_{-0.020}^{+0.021}$ & -1.864 \\
 13  &  5  & 3431 & 0.509 $\pm$ 0.007 & $0.963_{-0.029}^{+0.027}$ & -1.492 \\
 14  &  5  & 1712 & 0.525 $\pm$ 0.010 & $0.904_{-0.037}^{+0.035}$ & -2.814 \\
 15  &  5  & 879 & 0.541 $\pm$ 0.013 & $0.847_{-0.046}^{+0.045}$ & -3.297 \\
 16  &  5  & 355 & 0.569 $\pm$ 0.021 & $0.758_{-0.064}^{+0.065}$ & -3.473 \\
 17  &  5  & 201 & 0.556 $\pm$ 0.029 & $0.798_{-0.092}^{+0.090}$ & -2.141 \\
 18  &  5  &  95 & 0.520 $\pm$ 0.048 & $0.924_{-0.180}^{+0.166}$ & -0.516 ?\\

\hline\hline
  \end{tabular}
&
  \begin{tabular}{cccccc}
$R_{Void}^{Min}$ & $SW$ & $N$ & $\meanabscostheta$ & $p$ & SNR
\\
\hline\hline
 10  &  6  & 20068 & 0.501 $\pm$ 0.003 & $0.998_{-0.012}^{+0.012}$ & -0.193 \\
 11  &  6  & 12997 & 0.502 $\pm$ 0.003 & $0.990_{-0.014}^{+0.015}$ & -0.750 \\
 12  &  6  & 7777 & 0.506 $\pm$ 0.004 & $0.977_{-0.017}^{+0.019}$ & -1.355 \\
 13  &  6  & 4538 & 0.504 $\pm$ 0.007 & $0.985_{-0.025}^{+0.025}$ & -0.702 \\
 14  &  6  & 2283 & 0.510 $\pm$ 0.009 & $0.962_{-0.034}^{+0.032}$ & -1.256 \\
 15  &  6  & 1173 & 0.530 $\pm$ 0.011 & $0.887_{-0.040}^{+0.041}$ & -2.746 \\
 16  &  6  & 488 & 0.556 $\pm$ 0.018 & $0.798_{-0.058}^{+0.057}$ & -3.332 \\
 17  &  6  & 274 & 0.539 $\pm$ 0.024 & $0.856_{-0.084}^{+0.082}$ & -1.727 \\
 18  &  6  & 129 & 0.536 $\pm$ 0.038 & $0.866_{-0.134}^{+0.131}$ & -1.094 \\
\hline
 10  &  7  & 25473 & 0.501 $\pm$ 0.002 & $0.996_{-0.010}^{+0.010}$ & -0.436 \\
 11  &  7  & 16489 & 0.504 $\pm$ 0.003 & $0.986_{-0.013}^{+0.013}$ & -1.249 \\
 12  &  7  & 9807 & 0.507 $\pm$ 0.004 & $0.971_{-0.016}^{+0.016}$ & -1.925 \\
 13  &  7  & 5740 & 0.505 $\pm$ 0.006 & $0.980_{-0.021}^{+0.021}$ & -1.016 \\
 14  &  7  & 2878 & 0.511 $\pm$ 0.008 & $0.958_{-0.029}^{+0.029}$ & -1.534 \\
 15  &  7  & 1471 & 0.526 $\pm$ 0.010 & $0.903_{-0.038}^{+0.037}$ & -2.628 \\
 16  &  7  & 614 & 0.547 $\pm$ 0.016 & $0.830_{-0.055}^{+0.054}$ & -3.092 \\
 17  &  7  & 339 & 0.534 $\pm$ 0.021 & $0.872_{-0.076}^{+0.076}$ & -1.684 \\
 18  &  7  & 161 & 0.529 $\pm$ 0.035 & $0.889_{-0.125}^{+0.121}$ & -1.002 \\
\hline
 10  &  8  & 31560 & 0.500 $\pm$ 0.002 & $0.998_{-0.009}^{+0.010}$ & -0.192 \\
 11  &  8  & 20428 & 0.503 $\pm$ 0.003 & $0.990_{-0.012}^{+0.011}$ & -0.991 \\
 12  &  8  & 12239 & 0.505 $\pm$ 0.003 & $0.979_{-0.014}^{+0.015}$ & -1.565 \\
 13  &  8  & 7129 & 0.505 $\pm$ 0.004 & $0.981_{-0.019}^{+0.020}$ & -1.074 \\
 14  &  8  & 3609 & 0.508 $\pm$ 0.007 & $0.969_{-0.027}^{+0.028}$ & -1.260 \\
 15  &  8  & 1833 & 0.519 $\pm$ 0.009 & $0.928_{-0.034}^{+0.035}$ & -2.157 \\
 16  &  8  & 756 & 0.533 $\pm$ 0.015 & $0.875_{-0.052}^{+0.051}$ & -2.468 \\
 17  &  8  & 423 & 0.521 $\pm$ 0.019 & $0.918_{-0.073}^{+0.071}$ & -1.180 \\
 18  &  8  & 203 & 0.505 $\pm$ 0.030 & $0.978_{-0.120}^{+0.120}$ & -0.209 \\
\hline
 10  &  9  & 38273 & 0.503 $\pm$ 0.002 & $0.989_{-0.009}^{+0.009}$ & -1.500 \\
 11  &  9  & 24848 & 0.502 $\pm$ 0.002 & $0.991_{-0.011}^{+0.010}$ & -0.991 \\
 12  &  9  & 14941 & 0.505 $\pm$ 0.003 & $0.981_{-0.014}^{+0.014}$ & -1.599 \\
 13  &  9  & 8684 & 0.504 $\pm$ 0.004 & $0.984_{-0.017}^{+0.016}$ & -1.038 \\
 14  &  9  & 4368 & 0.510 $\pm$ 0.007 & $0.961_{-0.024}^{+0.024}$ & -1.747 \\
 15  &  9  & 2203 & 0.519 $\pm$ 0.009 & $0.927_{-0.031}^{+0.032}$ & -2.373 \\
 16  &  9  & 904 & 0.522 $\pm$ 0.013 & $0.917_{-0.048}^{+0.050}$ & -1.736 \\
 17  &  9  & 519 & 0.511 $\pm$ 0.017 & $0.955_{-0.068}^{+0.068}$ & -0.703 \\
 18  &  9  & 247 & 0.493 $\pm$ 0.028 & $1.028_{-0.111}^{+0.110}$ & 0.287 \\
\hline
 10  & 10  & 45522 & 0.502 $\pm$ 0.002 & $0.991_{-0.007}^{+0.009}$ & -1.260 \\
 11  & 10  & 29653 & 0.502 $\pm$ 0.002 & $0.994_{-0.010}^{+0.010}$ & -0.703 \\
 12  & 10  & 17929 & 0.505 $\pm$ 0.003 & $0.981_{-0.013}^{+0.012}$ & -1.723 \\
 13  & 10  & 10406 & 0.504 $\pm$ 0.004 & $0.983_{-0.017}^{+0.015}$ & -1.195 \\
 14  & 10  & 5226 & 0.510 $\pm$ 0.006 & $0.961_{-0.022}^{+0.022}$ & -1.938 \\
 15  & 10  & 2585 & 0.517 $\pm$ 0.007 & $0.935_{-0.029}^{+0.030}$ & -2.305 \\
 16  & 10  & 1078 & 0.523 $\pm$ 0.012 & $0.911_{-0.045}^{+0.045}$ & -2.054 \\
 17  & 10  & 624 & 0.516 $\pm$ 0.016 & $0.939_{-0.059}^{+0.061}$ & -1.062 \\
 18  & 10  & 284 & 0.509 $\pm$ 0.025 & $0.963_{-0.098}^{+0.097}$ & -0.425 \\
\hline

\hline\hline
  \end{tabular}
\\
\end{tabular}
\caption{\small 
  Statistics for subsamples with different sizes of voids and
  shells. This table corresponds to the ``cumulative samples'' in
  which each sample contains galaxies around voids with radius larger than
  $R_{Void}^{Min}$ .
  $R_{Void}^{Min}$: Lower limit of the void's radius in each sample, in
  \hmpc. 
  $SW:$ Width of the shell, in \hmpc. 
  $\meanabscostheta:$ Average of the distribution of \abscostheta.
  $p$: Resulting value of the $p$ parameter of
  Equation~(\ref{eq:pcostheta}).
  $SNR:$ Theoretical signal to noise ratio computed with Equation~(\ref{eq:snr}).
  Samples
  with less than 100 galaxies have been flagged with a
  question mark nearby the $SNR$ to remark the low reliability of
  these results. Errors in $\meanabscostheta$ and $p$ are computed using formulae in
  Appendix~\ref{app:uncertainties_p} in combination with Monte Carlo
  simulations. See text for more details.}
  \label{tab:p_rvoid_shellwidth}
\end{sidewaystable*}

%%%%%%%%%%%%%%%%%%%%%%%%%%%%%%%%%%%%%%%%%%%%%%%%%%
% TABLE: Results, no overlapping, differential
%%%%%%%%%%%%%%%%%%%%%%%%%%%%%%%%%%%%%%%%%%%%%%%%%%

\begin{sidewaystable*}
\scriptsize
\footnotesize
  \centering
\begin{tabular}{ll}

  \begin{tabular}{cccccc}
$R_{Void}$ & $SW$ & N & $\meanabscostheta$ & $p$ & SNR
\\
\hline\hline
 10.5  &  1  & 823 & 0.509 $\pm$ 0.014 & $0.966_{-0.053}^{+0.055}$ & -0.657 \\
 11.5  &  1  & 613 & 0.500 $\pm$ 0.017 & $1.001_{-0.066}^{+0.068}$ & 0.022 \\
 12.5  &  1  & 381 & 0.497 $\pm$ 0.020 & $1.014_{-0.084}^{+0.082}$ & 0.179 \\
 13.5  &  1  & 260 & 0.492 $\pm$ 0.025 & $1.031_{-0.105}^{+0.107}$ & 0.330 \\
 14.5  &  1  & 115 & 0.503 $\pm$ 0.038 & $0.987_{-0.153}^{+0.152}$ & -0.092 \\
 15.5  &  1  &  80 & 0.482 $\pm$ 0.044 & $1.076_{-0.195}^{+0.192}$ & 0.441 ?\\
 16.5  &  1  &  24 & 0.528 $\pm$ 0.080 & $0.893_{-0.328}^{+0.230}$ & -0.372 ?\\
 17.5  &  1  &  12 & 0.636 $\pm$ 0.104 & $0.572_{-0.316}^{+0.198}$ & -1.264 ?\\
 18.5  &  1  &  12 & 0.475 $\pm$ 0.145 & $1.105_{-1.166}^{+0.043}$ & 0.232 ?\\
\hline
 10.5  &  2  & 1685 & 0.501 $\pm$ 0.010 & $0.996_{-0.039}^{+0.039}$ & -0.114 \\
 11.5  &  2  & 1217 & 0.497 $\pm$ 0.012 & $1.011_{-0.048}^{+0.049}$ & 0.265 \\
 12.5  &  2  & 766 & 0.502 $\pm$ 0.014 & $0.990_{-0.058}^{+0.059}$ & -0.181 \\
 13.5  &  2  & 517 & 0.501 $\pm$ 0.018 & $0.996_{-0.073}^{+0.075}$ & -0.055 \\
 14.5  &  2  & 236 & 0.498 $\pm$ 0.027 & $1.009_{-0.112}^{+0.114}$ & 0.092 \\
 15.5  &  2  & 153 & 0.504 $\pm$ 0.032 & $0.986_{-0.131}^{+0.126}$ & -0.121 \\
 16.5  &  2  &  48 & 0.587 $\pm$ 0.054 & $0.702_{-0.165}^{+0.151}$ & -1.627 ?\\
 17.5  &  2  &  36 & 0.664 $\pm$ 0.057 & $0.506_{-0.133}^{+0.124}$ & -2.638 ?\\
 18.5  &  2  &  25 & 0.554 $\pm$ 0.118 & $0.805_{-0.622}^{+0.055}$ & -0.725 ?\\
\hline
 10.5  &  3  & 2715 & 0.494 $\pm$ 0.008 & $1.025_{-0.032}^{+0.031}$ & 0.851 \\
 11.5  &  3  & 1989 & 0.495 $\pm$ 0.009 & $1.019_{-0.038}^{+0.036}$ & 0.550 \\
 12.5  &  3  & 1225 & 0.506 $\pm$ 0.012 & $0.978_{-0.045}^{+0.046}$ & -0.532 \\
 13.5  &  3  & 837 & 0.500 $\pm$ 0.014 & $1.002_{-0.056}^{+0.057}$ & 0.030 \\
 14.5  &  3  & 397 & 0.517 $\pm$ 0.021 & $0.935_{-0.077}^{+0.079}$ & -0.902 \\
 15.5  &  3  & 250 & 0.511 $\pm$ 0.026 & $0.955_{-0.098}^{+0.100}$ & -0.485 \\
 16.5  &  3  &  77 & 0.597 $\pm$ 0.042 & $0.676_{-0.125}^{+0.116}$ & -2.278 ?\\
 17.5  &  3  &  57 & 0.620 $\pm$ 0.047 & $0.613_{-0.128}^{+0.120}$ & -2.427 ?\\
 18.5  &  3  &  45 & 0.574 $\pm$ 0.084 & $0.742_{-0.332}^{+0.143}$ & -1.331 ?\\
\hline
 10.5  &  4  & 3977 & 0.494 $\pm$ 0.006 & $1.023_{-0.027}^{+0.026}$ & 0.951 \\
 11.5  &  4  & 2851 & 0.497 $\pm$ 0.008 & $1.011_{-0.031}^{+0.031}$ & 0.377 \\
 12.5  &  4  & 1769 & 0.503 $\pm$ 0.010 & $0.988_{-0.039}^{+0.038}$ & -0.349 \\
 13.5  &  4  & 1234 & 0.485 $\pm$ 0.012 & $1.063_{-0.050}^{+0.050}$ & 1.440 \\
 14.5  &  4  & 597 & 0.515 $\pm$ 0.017 & $0.943_{-0.065}^{+0.064}$ & -0.965 \\
 15.5  &  4  & 376 & 0.518 $\pm$ 0.020 & $0.929_{-0.077}^{+0.076}$ & -0.962 \\
 16.5  &  4  & 108 & 0.596 $\pm$ 0.037 & $0.678_{-0.107}^{+0.104}$ & -2.671 \\
 17.5  &  4  &  78 & 0.594 $\pm$ 0.041 & $0.682_{-0.120}^{+0.116}$ & -2.239 ?\\
 18.5  &  4  &  70 & 0.527 $\pm$ 0.059 & $0.898_{-0.230}^{+0.184}$ & -0.603 ?\\
\hline
 10.5  &  5  & 5444 & 0.495 $\pm$ 0.006 & $1.019_{-0.024}^{+0.021}$ & 0.919 \\
 11.5  &  5  & 3980 & 0.497 $\pm$ 0.007 & $1.013_{-0.028}^{+0.027}$ & 0.536 \\
 12.5  &  5  & 2434 & 0.506 $\pm$ 0.009 & $0.977_{-0.032}^{+0.032}$ & -0.782 \\
 13.5  &  5  & 1719 & 0.491 $\pm$ 0.010 & $1.038_{-0.041}^{+0.041}$ & 1.041 \\
 14.5  &  5  & 833 & 0.507 $\pm$ 0.015 & $0.974_{-0.058}^{+0.057}$ & -0.515 \\
 15.5  &  5  & 524 & 0.520 $\pm$ 0.017 & $0.922_{-0.064}^{+0.065}$ & -1.238 \\
 16.5  &  5  & 154 & 0.583 $\pm$ 0.032 & $0.716_{-0.092}^{+0.092}$ & -2.752 \\
 17.5  &  5  & 106 & 0.581 $\pm$ 0.035 & $0.720_{-0.104}^{+0.100}$ & -2.249 \\
 18.5  &  5  &  95 & 0.520 $\pm$ 0.047 & $0.924_{-0.176}^{+0.171}$ & -0.516 ?\\

\hline\hline
  \end{tabular}

&

  \begin{tabular}{cccccc}
$R_{Void}$ & $SW$ & N & \abscostheta & $p$ & SNR
\\
\hline\hline
 10.5  &  6  & 7071 & 0.497 $\pm$ 0.004 & $1.014_{-0.020}^{+0.020}$ & 0.774 \\
 11.5  &  6  & 5220 & 0.500 $\pm$ 0.006 & $1.000_{-0.024}^{+0.024}$ & -0.013 \\
 12.5  &  6  & 3239 & 0.510 $\pm$ 0.008 & $0.961_{-0.027}^{+0.028}$ & -1.509 \\
 13.5  &  6  & 2255 & 0.496 $\pm$ 0.009 & $1.018_{-0.036}^{+0.036}$ & 0.555 \\
 14.5  &  6  & 1110 & 0.489 $\pm$ 0.013 & $1.046_{-0.056}^{+0.055}$ & 1.006 \\
 15.5  &  6  & 685 & 0.509 $\pm$ 0.015 & $0.964_{-0.059}^{+0.059}$ & -0.645 \\
 16.5  &  6  & 214 & 0.575 $\pm$ 0.027 & $0.738_{-0.083}^{+0.082}$ & -2.960 \\
 17.5  &  6  & 145 & 0.542 $\pm$ 0.031 & $0.845_{-0.105}^{+0.105}$ & -1.354 \\
 18.5  &  6  & 129 & 0.536 $\pm$ 0.038 & $0.866_{-0.134}^{+0.131}$ & -1.094 \\
\hline
 10.5  &  7  & 8984 & 0.496 $\pm$ 0.004 & $1.017_{-0.018}^{+0.018}$ & 1.044 \\
 11.5  &  7  & 6682 & 0.501 $\pm$ 0.006 & $0.997_{-0.022}^{+0.020}$ & -0.178 \\
 12.5  &  7  & 4067 & 0.512 $\pm$ 0.007 & $0.953_{-0.024}^{+0.025}$ & -2.051 \\
 13.5  &  7  & 2862 & 0.497 $\pm$ 0.008 & $1.012_{-0.032}^{+0.033}$ & 0.434 \\
 14.5  &  7  & 1407 & 0.494 $\pm$ 0.011 & $1.023_{-0.047}^{+0.046}$ & 0.560 \\
 15.5  &  7  & 857 & 0.511 $\pm$ 0.014 & $0.957_{-0.052}^{+0.052}$ & -0.859 \\
 16.5  &  7  & 275 & 0.562 $\pm$ 0.025 & $0.779_{-0.078}^{+0.078}$ & -2.763 \\
 17.5  &  7  & 178 & 0.538 $\pm$ 0.028 & $0.859_{-0.097}^{+0.096}$ & -1.356 \\
 18.5  &  7  & 161 & 0.529 $\pm$ 0.034 & $0.889_{-0.121}^{+0.119}$ & -1.002 \\
\hline
 10.5  &  8  & 11132 & 0.499 $\pm$ 0.004 & $1.006_{-0.016}^{+0.015}$ & 0.421 \\
 11.5  &  8  & 8189 & 0.499 $\pm$ 0.005 & $1.006_{-0.020}^{+0.017}$ & 0.348 \\
 12.5  &  8  & 5110 & 0.506 $\pm$ 0.005 & $0.976_{-0.022}^{+0.023}$ & -1.153 \\
 13.5  &  8  & 3520 & 0.501 $\pm$ 0.007 & $0.995_{-0.028}^{+0.028}$ & -0.202 \\
 14.5  &  8  & 1776 & 0.497 $\pm$ 0.010 & $1.011_{-0.040}^{+0.042}$ & 0.314 \\
 15.5  &  8  & 1077 & 0.508 $\pm$ 0.012 & $0.970_{-0.046}^{+0.047}$ & -0.674 \\
 16.5  &  8  & 333 & 0.549 $\pm$ 0.023 & $0.820_{-0.074}^{+0.075}$ & -2.417 \\
 17.5  &  8  & 220 & 0.533 $\pm$ 0.025 & $0.877_{-0.090}^{+0.087}$ & -1.300 \\
 18.5  &  8  & 203 & 0.505 $\pm$ 0.030 & $0.978_{-0.116}^{+0.119}$ & -0.209 \\
\hline
 10.5  &  9  & 13425 & 0.499 $\pm$ 0.003 & $1.005_{-0.015}^{+0.015}$ & 0.378 \\
 11.5  &  9  & 9907 & 0.499 $\pm$ 0.004 & $1.006_{-0.018}^{+0.016}$ & 0.394 \\
 12.5  &  9  & 6257 & 0.508 $\pm$ 0.006 & $0.967_{-0.021}^{+0.019}$ & -1.795 \\
 13.5  &  9  & 4316 & 0.499 $\pm$ 0.007 & $1.006_{-0.026}^{+0.025}$ & 0.250 \\
 14.5  &  9  & 2165 & 0.498 $\pm$ 0.009 & $1.008_{-0.037}^{+0.039}$ & 0.243 \\
 15.5  &  9  & 1299 & 0.519 $\pm$ 0.011 & $0.927_{-0.041}^{+0.041}$ & -1.840 \\
 16.5  &  9  & 385 & 0.533 $\pm$ 0.021 & $0.877_{-0.073}^{+0.074}$ & -1.720 \\
 17.5  &  9  & 272 & 0.525 $\pm$ 0.024 & $0.904_{-0.087}^{+0.085}$ & -1.120 \\
 18.5  &  9  & 247 & 0.493 $\pm$ 0.027 & $1.028_{-0.110}^{+0.110}$ & 0.287 \\
\hline
 10.5  & 10  & 15869 & 0.503 $\pm$ 0.003 & $0.986_{-0.012}^{+0.013}$ & -1.173 \\
 11.5  & 10  & 11724 & 0.497 $\pm$ 0.003 & $1.014_{-0.016}^{+0.016}$ & 1.013 \\
 12.5  & 10  & 7523 & 0.508 $\pm$ 0.004 & $0.969_{-0.019}^{+0.019}$ & -1.852 \\
 13.5  & 10  & 5180 & 0.499 $\pm$ 0.006 & $1.005_{-0.024}^{+0.023}$ & 0.220 \\
 14.5  & 10  & 2641 & 0.503 $\pm$ 0.008 & $0.988_{-0.034}^{+0.033}$ & -0.424 \\
 15.5  & 10  & 1507 & 0.513 $\pm$ 0.011 & $0.950_{-0.039}^{+0.041}$ & -1.324 \\
 16.5  & 10  & 454 & 0.536 $\pm$ 0.020 & $0.867_{-0.068}^{+0.067}$ & -2.035 \\
 17.5  & 10  & 340 & 0.521 $\pm$ 0.021 & $0.921_{-0.079}^{+0.078}$ & -1.023 \\
 18.5  & 10  & 284 & 0.509 $\pm$ 0.025 & $0.963_{-0.096}^{+0.094}$ & -0.425 \\
\hline

\hline\hline
  \end{tabular}

\\
\end{tabular}

\caption{\small Statistics for subsamples with different size of voids
  and shells. This table corresponds to the ``differential samples''
  in which each sample contains galaxies around voids with radii in
  the intervals ($R_{Void}-0.5,R_{Void}+0.5)$.
  $R_{Void}$: Midpoint of each interval, in \hmpc.
  $SW:$ Width of the shell, in \hmpc.  $\meanabscostheta:$ Average of
  the distribution of \abscostheta.  $p$: Resulting value of the $p$
  parameter of Equation~(\ref{eq:pcostheta}).  $SNR:$ Theoretical
  signal to noise ratio computed with Equation~(\ref{eq:snr}). Samples
  with less than 100 galaxies have been flagged with a
  question mark nearby the $SNR$ to remark the low reliability of
  these results. Errors
  in $\meanabscostheta$ and $p$ are computed using formulae in
  Appendix~\ref{app:uncertainties_p} in combination with Monte Carlo
  simulations. See text for more details.}
  \label{tab:p_rvoid_shellwidth_dif}
\end{sidewaystable*}

%%%%%%%%%%%%%%%%%%%%%%%%%%%%%%%%%%%%%%%%%%%%%%%%%%
% FIGURE: Results, No overlapping, cumulative
%%%%%%%%%%%%%%%%%%%%%%%%%%%%%%%%%%%%%%%%%%%%%%%%%%

%%%%%%%%%%%%%%%%%%%%%%%%%%%%%%%%%%%%%%%%%%%%%%%%%%
% TABLE: Rvoid Cumulative, SW<5, SW>5 Mpc
%%%%%%%%%%%%%%%%%%%%%%%%%%%%%%%%%%%%%%%%%%%%%%%%%%

\begin{table*}
  \centering
  \begin{tabular}[bt]{c|cccc|cccc}
$R_{Void}^{Min}$  & \multicolumn{4}{c}{$0\,\hmpc<R_{Shell}\leq5\,\hmpc$} & 
\multicolumn{4}{c}{$5\,\hmpc<R_{Shell}\leq10\,\hmpc$} \\
\cline{2-5}\cline{6-9}
& $N$ & $\meanabscostheta$ & $p$ & SNR &
 $N$ & $\meanabscostheta$ & $p$ & SNR\\
\hline\hline
10  &  15289  &  0.500  $\pm$  0.003  & $ 0.999 _{-0.012}^{+0.012 }$ &  -0.053  &  30233  &  0.502  $\pm$  0.002  & $ 0.992_{-0.008}^{+0.009 }$ &  -0.915 \\
11  &  9845  &  0.505  $\pm$  0.004  & $ 0.980 _{-0.015}^{+0.015 }$ &  -1.365  &  19808  &  0.501  $\pm$  0.003  & $ 0.996_{-0.010}^{+0.011 }$ &  -0.374 \\
12  &  5865  &  0.509  $\pm$  0.005  & $ 0.964 _{-0.018}^{+0.019 }$ &  -1.864  &  12064  &  0.504  $\pm$  0.003  & $ 0.984_{-0.013}^{+0.014 }$ &  -1.170 \\
13  &  3431  &  0.509  $\pm$  0.006  & $ 0.963 _{-0.025}^{+0.025 }$ &  -1.492  &  6975  &  0.503  $\pm$  0.004  & $ 0.988_{-0.018}^{+0.017 }$ &  -0.698 \\
14  &  1712  &  0.525  $\pm$  0.009  & $ 0.904 _{-0.033}^{+0.033 }$ &  -2.814  &  3514  &  0.502  $\pm$  0.006  & $ 0.992_{-0.024}^{+0.026 }$ &  -0.304 \\
15  &  879  &  0.541  $\pm$  0.013  & $ 0.847 _{-0.043}^{+0.044 }$ &  -3.297  &  1706  &  0.504  $\pm$  0.009  & $ 0.985_{-0.035}^{+0.036 }$ &  -0.428 \\
16  &  355  &  0.569  $\pm$  0.020  & $ 0.758 _{-0.060}^{+0.065 }$ &  -3.473  &  723  &  0.495  $\pm$  0.014  & $ 1.020_{-0.055}^{+0.058 }$ &  0.350 \\
17  &  201  &  0.556  $\pm$  0.027  & $ 0.798 _{-0.083}^{+0.090 }$ &  -2.141  &  423  &  0.491  $\pm$  0.018  & $ 1.037_{-0.073}^{+0.077 }$ &  0.496 \\
18  &  95  &  0.520  $\pm$  0.039  & $ 0.924 _{-0.133}^{+0.154 }$ &  -0.516  &  189  &  0.502  $\pm$  0.027  & $ 0.991_{-0.102}^{+0.114 }$ &  -0.080 \\

\hline\hline
  \end{tabular}
  \caption{Strength of the alignment for galaxies within a shell up to
    $5\,\hmpc$ (left) and for galaxies within a shell between $5\,\hmpc$ and
    $10\,\hmpc$ (right). Each line corresponds to samples of $N$
    galaxies around voids with radius larger than
    $R^{Min}_{Void}$. Errors are computed using theoretical
    expressions in Appendix~\ref{app:uncertainties_p}.}
  \label{tab:signal_two_shells}
\end{table*}

%%%%%%%%%%%%%%%%%%%%%%%%%%%%%%%%%%%%%%%%%%%%%%%%%
% FIGURE: Results, No overlapping, differential
%%%%%%%%%%%%%%%%%%%%%%%%%%%%%%%%%%%%%%%%%%%%%%%%%%

%%%%%%%%%%%%%%%%%%%%%%%%%%%%%%%%%%%%%%%%%%%%%%%%%%
% FIGURE: SNR vs Ngal
%%%%%%%%%%%%%%%%%%%%%%%%%%%%%%%%%%%%%%%%%%%%%%%%%%

%%%%%%%%%%%%%%%%%%%%%%%%%%%%%%%%%%%%%%%%%%%%%%%%%%
% FIGURE: Histogram of cos(theta) for better SNR
%%%%%%%%%%%%%%%%%%%%%%%%%%%%%%%%%%%%%%%%%%%%%%%%%%
\def\scl{0.4}
\begin{figure*}
  \centering
  \includegraphics[scale=\scl]{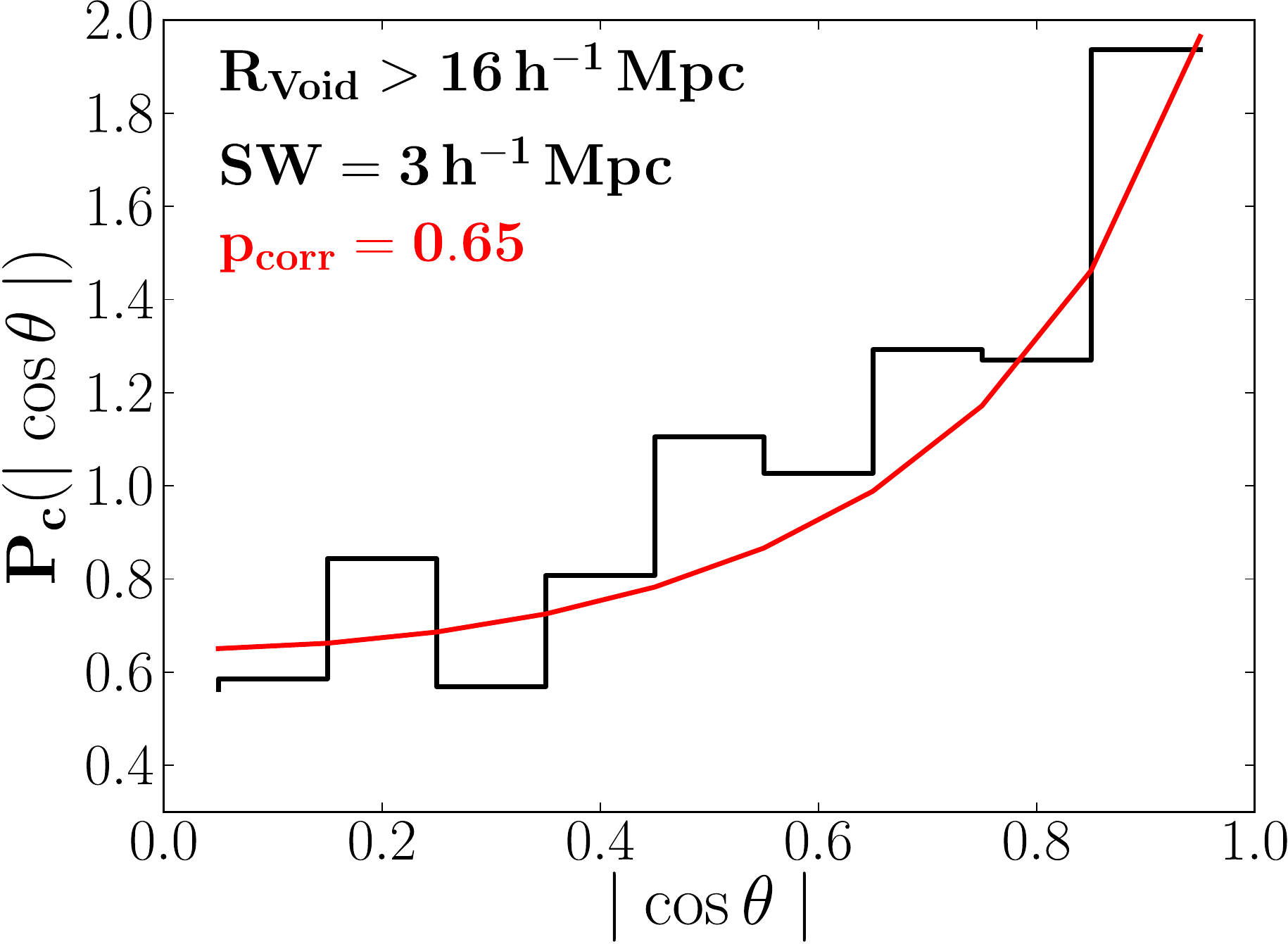}
  \includegraphics[scale=\scl]{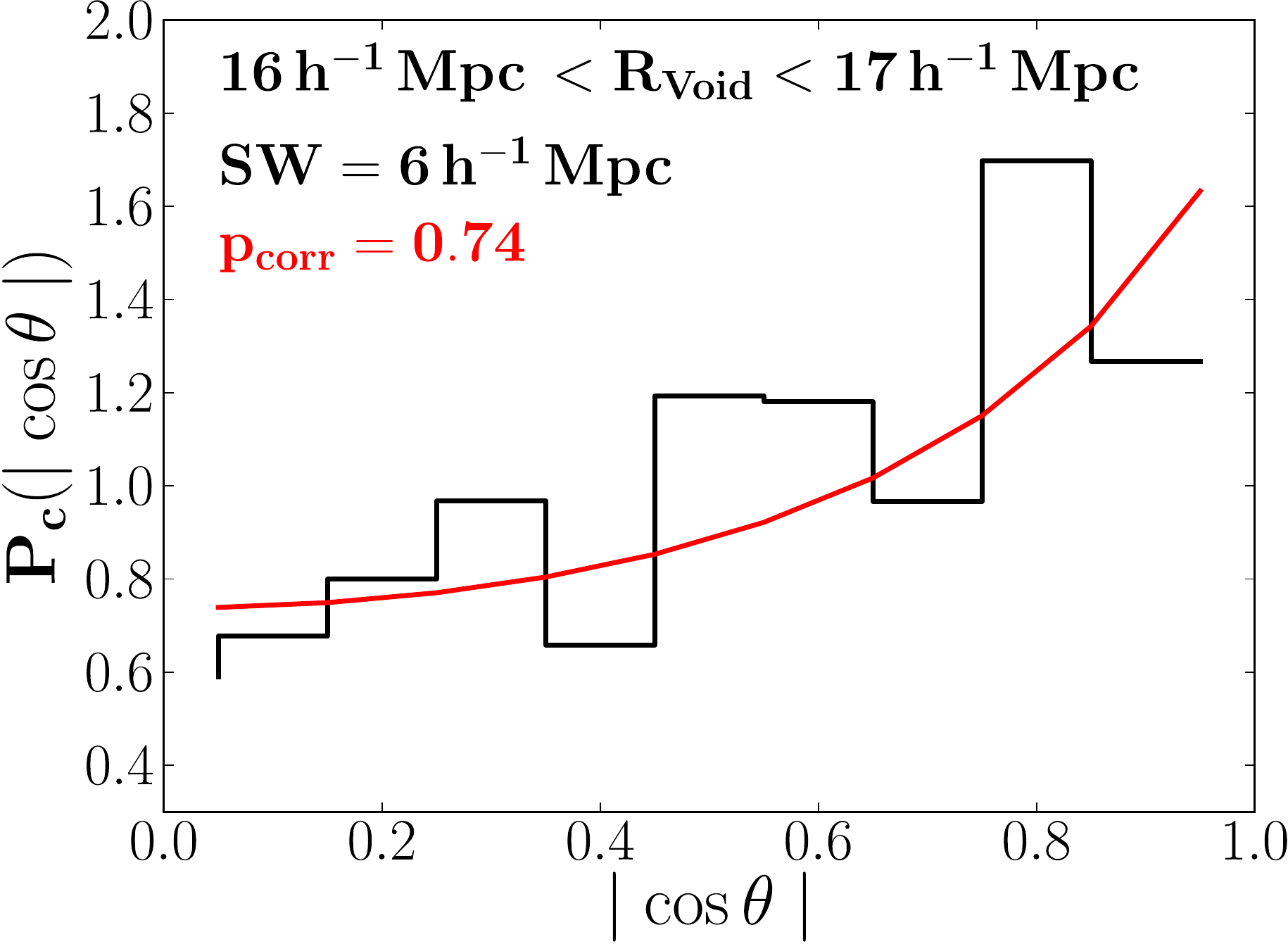}
  \caption{Distribution of \abscostheta\  after applying the statistical
    correction for the two subsamples that reach the highest $|SNR|$
    in Table~\ref{tab:p_rvoid_shellwidth} (left) and 
    Table~\ref{tab:p_rvoid_shellwidth_dif} (right). The continuous
    line corresponds to the theoretical model described by
    Equation~(\ref{eq:pcostheta}) for the measured value of $p$, shown
    in the upper left corners.
    \label{fig:Pcorr_hist}}
\end{figure*}

%%%%%%%%%%%%%%%%%%%%%%%%%%%%%%%%%%%%%%%%%%%%%%%%%%%%%%%%%%%%%%%%%%%%%%
% 
% Discussion
%
%%%%%%%%%%%%%%%%%%%%%%%%%%%%%%%%%%%%%%%%%%%%%%%%%%%%%%%%%%%%%%%%%%%%%%

\section{Discussion\label{sec:discussion}}

We have analysed a large sample of galaxies around 699 voids with
radius larger than $10\,\hmpc$ up to $z=0.12$. We have found that for
voids with radius $\gtrsim15\,\hmpc$ and within a shell not larger
than $\sim 5\,\hmpc$, disk galaxies present a significant
tendency to have their spin vectors aligned with the
radial direction of the void.

The maximum $|SNR|=3.62$ is measured for voids with $R_{Void}\geq
16\,\hmpc$ and a shell width of $3\,\hmpc$ with a strength of the
alignment $p=0.664^{+0.083}_{-0.074}$. However, this value gives an
overestimation of the real strength since has been selected as the
best case out of many subsamples.

In the next sections we compare our results with previous empirical
works and with results from numerical simulations.

\subsection{Comparison with empirical works}

From the observational point of view,~\citetalias{Trujillo2006} and,
more recently,~\citetalias{Slosar2009} have performed a similar
analysis to the one done here. \citetalias{Trujillo2006} analysed 201
face-on and edge-on galaxies around voids with $R>10\,\hmpc$ using
data from the SDSS-DR3 and the 2dFRGS. They found a 
significant tendency of the spin of the galaxies to be in the
direction perpendicular to the radial direction of the void. More
recently, \citetalias{Slosar2009} using two samples of 578 and 258
galaxies from the SDSS-DR6 with similar selection criteria found no
statistical evidence for departure from random orientations.

Using the same criteria on the size of the voids
($R_{Void}>10\,\hmpc$) and on the width of the shell ($4\,\hmpc$) as
in those previous works, we find no significance alignment ($p=0.998$;
$SNR=-0.15$; see Table~\ref{tab:p_rvoid_shellwidth}). The size of the
sample used to establish this result is of 11060 galaxies, which after
applying the correction factor of 0.6, means an equivalent size of
6636 galaxies. This number is 8 times larger the size used by
\citetalias{Slosar2009}.

For a better comparison, we have computed the signal of the alignment
using criteria similar to that of \citetalias{Slosar2009} regarding
the selection of spiral galaxies ($g-r<0.6$) and the definition of
edge-on ($b/a<0.27$) and face-on ($b/a<0.96$) galaxies\footnote{We
  used an slightly different way to compute the axial ratio $b/a$
  compared with \citetalias{Slosar2009}, however, we do not consider
  this to have a significance effect on our comparison.}. However, we
keep our limit in $M_r$ instead of using $M_r>-21+5\log h$ as done by
\citetalias{Slosar2009} because otherwise the final number of galaxies
would be too small. After applying these criteria, we finished with a
sample of 252 face-on and edge-on spiral galaxies. The value of $p$
obtained with this sample is 0.993 with a $SNR=0.1$ and, therefore,
compatible with a random distribution.

In relation to \citetalias{Trujillo2006}, it is important
  to note that in that work, the signal presented corresponds to the
  peak of the signal found after exploring in shells of different
  widths. Consequently, to properly address the signification of the
  signal in \citetalias{Trujillo2006} we have reviewed the data used
  in that paper taking into account this exploration. To do this, we
  have run simulations with similar number of galaxies around similar
  voids ($R>10\,\hmpc$) searching for the maximum of the $|SNR|$ in
  shells of width from 3\,\hmpc\ to 7\,\hmpc\ in steps of
  0.1\,\hmpc. Then, it has been computed the fraction of simulations
  with $\max(|SNR|)$ larger than the one found in \citetalias{Trujillo2006}
  when only SDSS data was used ($\max(SNR_{T06;SDSS})=2$)\footnote{In the
    process of conducting this analysis we noted some duplications in
    a few galaxies that make the measured $SNR$ decreases from 2.4 to
    2.}. After doing our analysis we found that $\sim16\%$ of the
  simulations showed $\max(|SNR|)>2$, decreasing the significance of
  the result of \citetalias{Trujillo2006} to $\sim84\%$.

Another observational work on alignment of galaxies with respect to
the local large scale structure is that by~\citet{Lee2007a}. In this
paper, the authors computed the shear tidal tensor in the position of
each galaxy and measured the angle between each of the principal axes
and the spin direction of the galaxy. They found a significant
alignment ($>6\sigma$) between the direction of the spin and the
intermediate principal axis of the shear tidal tensor.\footnote{The
  authors obtained a value of $c=0.084\pm0.014$ which corresponds to
  $p=1.057\pm 0.012$} Nevertheless, the comparison with our results is
not direct since we do not use a direct measure of the orientation of
the shear tidal tensor in the position of each galaxy and the radial
direction of the voids can be considered only as a statistical proxy
for the direction of the major principal axis of the shear tidal
tensor. Given the differences in methodology, a meaningful comparison
of the results from both works would need an analysis that it is
beyond the scope of this paper.

\subsection{Comparison with numerical simulations}

Another way to study the alignment of galaxies with their local large
scale environment is through numerical simulations
\citep{Porciani2002a, Porciani2002b, Navarro2004,
    Bailin2005a, Bailin2005b, Altay2006, Patiri2006c, Brunino2007, Aragon-Calvo2007,Cuesta2008a,
    Zhang2009, Hahn2010, Bett2010}. Since the behaviour
  of the halos can be dependent on the environment or the large scale
  structure in which they reside, to perform a meaningful comparison
  we have focused on  the analysis done by \citet{Patiri2006c},
\citet{Brunino2007} and \citet{Cuesta2008a} in which it was studied
the orientation of dark matter halos around cosmic voids using
different cosmological simulations. The criteria imposed to the dark
matter halos and the procedure to detect voids tried to match the
criteria used in~\citetalias{Trujillo2006}. All three works found that
the minor axis and the major axis of the halos have significant
tendencies to be aligned with the radial and the perpendicular
directions, respectively. The results regarding the orientation of the
angular momentum of the halos were less clear. \citet{Patiri2006c} did
not find any particular orientation for the angular momentum of the
halos. \citet{Brunino2007} also did not found any particular alignment
for the angular momentum in their full sample of halos although they
detected a tendency for those halos with a disc-dominated galaxy to
have their angular momentum perpendicular to the radial
direction. Finally, \citet{Cuesta2008a} measured a significant
($>7\sigma$) tendency of the spin of the dark matter halos to lie in
the plane perpendicular to the radial direction. However, the same
authors found that the strength of the alignment is mainly produced by
the outer regions of the DM halos and this would explain the
discrepancies with \citet{Brunino2007} were the inner regions of the
DM were used to measure the alignment.

How these results relate with ours is not straightforward since we
observe the collapsed baryonic matter and they studied the dark matter
or the non-collapsed baryonic matter. We can only point out the fact that
the alignment that we find in the galaxies is shared by the minor axis
of the halos studied in the simulations, either dark matter or gas
halos. This is suggestive to an interaction between the galaxy and the
hosting halos around it (either of dark or baryonic matter) leading to
a tendency of the minor axis of the galaxy (and, therefore, its
angular momentum) to be aligned with the minor axis of the halo's
matter distribution.

\section{Summary\label{sec:summary}}

Analysing a volume of $\sim 27\times10^6\,(\hmpc)^3$ from the SDSS-DR7
we have searched for cosmic voids devoid of galaxies brighter than
$M_r-5\log h=-20.17$ and with $R_{Void}>10\,\hmpc$. We have found 699
non overlapping voids for which we provide positions and sizes.

We have used this catalog of voids to search for disk galaxies
around them and study the alignment between the direction of the
angular momentum of these galaxies and the radial direction with
respect to the center of the voids. 

We have included two improvements with respect to previous similar
works. 

First, we have used an updated version of the SDSS
spectroscopic catalog (data release 7) and we have combined it with
the visual morphological classification from the Galaxy Zoo project to
get a reliable sample of disk galaxies.

Second and more important, we have introduced a statistical procedure
that has allowed us to overcome the problem of the indetermination of
the real inclination of galaxies computed from their apparent axial
ratio. We have performed extensive Monte Carlo simulations to check
the validity of this procedure. We show that the procedure recover the
real signal without practically any bias and its power in terms of
capacity to reject the null hypothesis it is equivalent to the case in
which it is used a sample with complete knowledge of the real
direction of the spin of the galaxies using $60\%$ the amount of
galaxies. In comparison with the common procedure of selecting only
edge-on and face-on galaxies, this procedure means an increase of
about a factor 3 in the amount of measurements used in the analysis of
the alignment.

These improvements have allowed us to detect a statistically
significant ($\gtrsim98.8\%$) tendency of galaxies around very large voids
($R\gtrsim\,15\hmpc$) to have their angular momentum align with the
radial direction of the voids. However, for smaller voids this
tendency disappears and the results are consistent with no special
alignment.

We have also found that the strength of the alignment depends on the
distance of the galaxies to the surface of the voids and for galaxies
further than \mbox{$\sim5\,\hmpc$} the distribution of the alignments is
compatible with a random distribution independent of the size of the
voids.

Previous similar works found opposite alignment
\citepalias{Trujillo2006} or no alignment
\citepalias{Slosar2009}. However, these works used too few galaxies
around voids with $R\geq10\,\hmpc$ which, according to our work, could
mask the signal. In fact, using the same criteria for the size of the
voids and the width of the shells as in those works, our data is
compatible with a random distribution of spins without any particular
alignment, as found by~\citetalias{Slosar2009}.

The comparison with the results from cosmological simulations points
to a possible connection between the alignment of the halos (of dark
matter and non-collapse baryonic matter) and that of the galaxies
which could explain the similar orientation of both components
observed in the simulations and in our work, respectively.

%%%%%%%%%%%%%%%%%%%%%%%%%%%%%%%%%%%%%%%%%%%%%%%%%%%%%%%%%%%%%%%%%%%%%%
%
% Bibliography
%
%%%%%%%%%%%%%%%%%%%%%%%%%%%%%%%%%%%%%%%%%%%%%%%%%%%%%%%%%%%%%%%%%%%%%%

\bibliographystyle{aa}
\bibliography{Bibliography}

%%%%%%%%%%%%%%%%%%%%%%%%%%%%%%%%%%%%%%%%%%%%%%%%%%%%%%%%%%%%%%%%%%%%%%
%
% Appendices
%
%%%%%%%%%%%%%%%%%%%%%%%%%%%%%%%%%%%%%%%%%%%%%%%%%%%%%%%%%%%%%%%%%%%%%%
\clearpage
\appendix

%%%%%%%%%%%%%%%%%%%%%%%%%%%%%%%%%%%%%%%%%%%%%%%%%%%%%%%%%%%%%%%%%%%%%%
% Statistical correction
%%%%%%%%%%%%%%%%%%%%%%%%%%%%%%%%%%%%%%%%%%%%%%%%%%%%%%%%%%%%%%%%%%%%%%
\section{Statistical computation of $P(\abscostheta)$ with full
  information\label{sec:appendixA}}

In this Appendix we describe in detail the statistical procedure that
has been used to compute the corrected value of $p$ using all the disk
galaxies, independently of their inclination. We also show its
validity and robustness using Monte Carlo simulations.

\subsection{Mathematical justification of the procedure}

In the analysis of the alignment of galaxies, one of the main sources
of uncertainty is the indetermination in the inclination of the plane
of a disk galaxy with respect to the line of sight $\pm\zeta$ using
exclusively the observed axial ratio $b/a$ of the galaxies (see
Equation~\ref{eq:inclination}). In other words, if a galaxy is
divided in two halves separated by its major axis, it is not possible
to know which of the two halves is the closest to the
observer.\footnote{The use of kinematic information or the presence of
  dust lanes can help to break this indetermination, however, in most
  of the cases this is information is not accessible.} This
indetermination is negligible for edge-on ($b/a\sim0$) and face-on
galaxies ($b/a\sim1$) but using only these galaxies reduces the sample
size to $\sim 1/5$ of the original one. As we increase the range of
allowed values of $b/a$, the increasing uncertainties in $\zeta$ will
result in an increasing degradation of any existing alignment but the
statistics improve. The question is whether this improvement of the
statistics can compensate for the increasing degradation of any
possible signal. The answer is ``yes''. Choosing always the plus sign
in the computation of $\zeta$ , or the minus sign, or any random
assignment of signs, leads to the same statistical results (i.e. they
are equally powerful tests), which are better than those obtained with
any limitation of the range of possible values of $b/a$. However, the
estimate of the alignment obtained in this manner is biased towards
smaller values (the strength of the alignment is given by $(1-p)\simeq
-3c/4$ for weak alignments). This would not be much of a problem,
because one may calibrate the procedure using Monte Carlo simulations
and then correct for the biasing. In this manner we have found (see
Table~\ref{tab:p_check}):
\begin{equation}
  1-p_+ \simeq 0.6(1-p)
\end{equation}

where $p$ corresponds to the real alignment and $p_+$ is the value
obtained using the plus sign for $\zeta$.

The main problem with the use of $p_+$, or any other sign assignment,
is that it introduces an artificial randomness that increases the
scatter of the estimates. Using $p_+$, we assign the correct sign to
half of the galaxies, on average, while the other half gets the wrong
sign, but the exact number of galaxies getting the correct sign
fluctuates from sample to sample (with variance $N_g/4$, being $N_g$
the size of the sample) resulting in an increased error. Furthermore,
since we do not take into account the other possible sign assignment,
we do not know how large is the degradation of the alignment implied
by those galaxies that get the wrong sign.

To avoid these problems we propose a method that uses all the
information in the data and does not introduce artificial
randomness. To this end, we consider the two possible values of
$\theta$ associated with every galaxy (one value for each possible
sign of $\zeta$) and assume that only half of the values of $\theta$
falling in a given range are correct while the other half is
incorrect. The correct values for the latter half of galaxies would be
the conjugate of $\theta$, $\theta'$, corresponding to the value of
$\theta$ using the opposite sign for $\zeta$. Thus, if the actual
probability distribution for $\theta$ were:\footnote{We are only
  interested in the direction of the alignment and therefore the
  analysis can be restricted to $0\leq \theta \leq \pi/2$ and $\cos
  \theta = \abscostheta$.}

\begin{equation}
\label{eq:p_cos_theta}
  \bar{P}(\cos \theta,p) = \frac{p}{(1+(p^2-1)\cos^2\theta)^{3/2})}\,,
\end{equation}

the probability distribution that would be inferred from the $2N_g$
values of $\theta$ treating them as if they were independent, $P(\cos \theta)$,
would be given, for the $j-th$ bin, by:
\begin{equation}
  P(\cos \theta_j) = \frac{1}{2}\bar{P}(\cos \theta_j,p) +
  \frac{1}{2l}\sum_{i=1}^l \bar{P}(\cos \theta_j',p),
\end{equation}

where $l$ is the number of $\theta_j$ values in the $j-th$ bin.

This formula expresses the fact that with probability 1/2, the
probability density in the bin centered in $\theta_j$ is given by the
real distribution $\bar{P}$ evaluated at $\theta_j$ (correct sign
assignment), while with probability 1/2, the probability density at
$\theta_j$ is the averaged of the value of $\bar{P}$ over the
conjugate values ($\theta_j'(i)$) of the $l$ values of $\theta$
falling in bin $j$.

So, the factor:
\begin{equation}
\label{eq:q_factor}
  Q(\cos \theta_j) \equiv \frac{\bar{P}(\cos \theta_j)}
{\frac{1}{2}\bar{P}(\cos
  \theta_j,p)+\frac{1}{2l}\sum_{i=1}^l\bar{P}(\cos \theta_j'(i),p)}
\end{equation}

is an estimate of the ratio between the actual
  distribution, $\bar{P}(\cos \theta_j)$, and the first estimate,
$P(\cos \theta_j)$.

Therefore, we have for the estimate of $\bar{P}$ (that we denote by $P_c$):
\begin{equation}
\label{eq:p_corr}
  P_c(\cos \theta_j) = P(\cos \theta_j) Q(\cos \theta_j)
\end{equation}

When the alignment is very strong, the assumption that the two values
$\theta, \theta'$ of a conjugate couple have the same probability can
no longer be mantained. Instead, we should used:

\begin{equation}
  \textrm{Prob}(\theta) = \frac{P(\cos \theta)}{P(\cos \theta)+P(\cos \theta')}
\end{equation}

\begin{equation}
  \textrm{Prob}(\theta') = \frac{P(\cos \theta')}{P(\cos \theta)+P(\cos \theta')}
\end{equation}

and modify the definition of $Q$ consequently. However, this
complication of the method is not worthy to our purpose. In fact, from
Table~\ref{tab:p_check}, we can see that even for considerable
alignment strengths, the bias implied by neglecting this last
refinement is small, and can be corrected by the following expression:
\begin{equation}
  p_{db}= 1-(1+0.1(1-p_c)^2)(1-p_c)
\end{equation}

where $p_c$ is the value obtained using Equation~(\ref{eq:p_corr}),
and $p_{db}$ is the debiased value.

From Table~\ref{tab:p_check} we can also see that the relative error,
$\sigma_p/|1-p|$, is always larger for $p_+$ or $p_-$ than for
$p_c$. For weak alignments ($|1-p|\lesssim0.1$) the former is
$\sim20\%$ larger than the latter, while for larger alignments the difference
diminishes.

Finally, it must be noticed that the method that we have just
described does not depend on the form of $\bar{P}(\cos \theta)$.

\subsection{Description of the procedure}

In this appendix we describe the actual implementation of the method
presented above.
 
The procedure is as follows:

\begin{enumerate}
\item For each galaxy, we compute the two possible values of
  \costheta\ corresponding to the two alternatives signs of $\zeta$
  and hence the two possible spin orientations.
\item Then, we construct a normalized histogram assuming the two
  values of \costheta\ of each galaxy as independent values. The
  normalization is done dividing each bin by 2 times the total number
  of galaxies of the sample ($N_g$) and by the width of the bins.  We
  call this non-corrected histogram $P(\cos \theta)$.

\item Next, in each bin centered in $\cos \theta_j$, we compute the
  value of the corrected histogram $P_c(\cos \theta)$ using
  Equations~(\ref{eq:p_corr}) and (\ref{eq:q_factor}):

  \begin{equation}
    \label{eq:pcorr}
    P_c(\cos \theta_j)  =  P(\cos \theta_j)\,Q(\cos \theta_j)
  \end{equation}

    remembering that

  \begin{eqnarray}
    Q(\cos \theta_j) &=& \frac{\bar{P}(\cos \theta_j,p)}{\frac{1}{2}\bar{P}(\cos
    \theta_j,p)+\frac{1}{2l}\sum_{i=1}^l \bar{P}(\cos
    \theta^\prime_j(i),p)} \nonumber
  \end{eqnarray}

  \normalsize

  and that $\theta'_j(i)$ are the conjugate values of $\theta$ for
  those galaxies with $\theta_j(i)$ within the interval \mbox{$\mid
    \theta_j(i)-\theta_j \mid \leq \Delta\theta_j/2$}, $l$ is the
  total number of values within the bin, and
  \begin{equation}
    \bar{P}(\cos \theta,p) \equiv \frac{p}{(1+(p^2-1)\cos^2\theta)^{3/2}}.
  \end{equation}
\end{enumerate}

The final corrected value of $p$ is computed numerically using its
relationship with $\langle \costheta \rangle$ from
Equation~(\ref{eq:p_from_cos}), which given the distribution $P_c(\cos
\theta_j)$ as a discrete distribution can be expressed as:

\begin{equation}
  \label{eq:corrected_p}
  \frac{\sum_{j=1}^n P_c(\cos \theta_j) \cos \theta_j}{\sum_{j=1}^n
    P_c(\cos \theta_j)} = \frac{1}{1+p}
\end{equation}

with $n$ the total number of bins in which the distribution is divided.

%%%%%%%%%%%%%%%%%%%%%%%%%%%%%%%%%%%%%%%%%%%%%%%%%%%%%%%%%%%%%%%%%%%%%%
% Robustness of the statistical correction
%%%%%%%%%%%%%%%%%%%%%%%%%%%%%%%%%%%%%%%%%%%%%%%%%%%%%%%%%%%%%%%%%%%%%%
\subsection{Robustness of the statistical correction\label{app:robustness}}

To check the robustness of the statistical correction we have
performed a series of Monte Carlo simulations. In these simulations,
we use samples of fake galaxies in the position of the real ones but
with spin directions assigned randomly following a $p-distribution$
(Equations~\ref{eq:p_cos_theta}) with a given $p_{input}$. Then, the
samples of fake galaxies are analysed in the same manner of the real
galaxies and a final $p_{output}$ value is obtained.

We have run 2 sets of simulations using two samples with different
number of galaxies ($N_g$) to check the robustness of the procedure
also as a function of the sample size. These samples correspond to
galaxies in shells of $4\,\hmpc$ and $R_{Void}>10\,\hmpc$ (Sample A,
following the usual criteria used in previous works) and to galaxies
in shells of $3\,\hmpc$ and $R_{Void}>16\,\hmpc$ (Sample B,
corresponding to our maximum $SNR$). For each sample we have run 1000
Monte Carlo realizations with 7 different initial distributions of
$\meanabscostheta$ described by their corresponding $p$
values ($p_{input}$). These values covered the typical values of $p$
that we have found in our analysis.

Table~\ref{tab:p_check} shows the results of this analysis. For each
subset of 1000 realizations we give the size of the sample $N$, the
input value $p_{input}$, the mean value of $p$ obtained if a fixed
sign for $\zeta$ is used ($p_+$ and $p_-$, for plus and minor sign,
respectively) and the mean value of $p$ when applying our stastical
correction, $p_{output}$. The uncertainties shown correspond to
$1\sigma$ of the distribution of the single values in the 1000
realizations.

We found that for most of the cases the statistically corrected value is
within $1\sigma$ of the input value showing the high accuracy of the
procedure, especially when comparing with the cases in which a fixed
sign is used.

The results of these simulations have been used to compute the
``effective size'' of the initial sample. This effective size is
defined as the size that a sample with complete knowledge of the real
signs of $\zeta$ for each galaxy should have to show the same
uncertainties that we find in our simulations. On what follows, it is
described how we have computed the correction factor to be applied to
our samples to obtain their effective sizes.

It can be proved theoretically that the value of the standard
deviation of $\meanabscostheta$, $\sigma_{\meanabscostheta}$, for the
case in which there is no preferential alignment ($p=1$), is:

\begin{equation}
  \label{eq:std_cos_theo}
  \sigma_{\meanabscostheta} = \frac{1}{\sqrt{12N}},
\end{equation}

where $N$ is the total number of galaxies used to compute
\meanabscostheta. However, this theoretical expression assumes the
full knowledge of the values of $\theta$ for all the galaxies, while
empirically we do not have such full information because of the
indetermination in the sign of $\zeta$. Therefore, we have compared
the standard deviation obtained from the simulations with different
values of $N$, with the theoretical value. From this comparison, we
have obtained a correction factor to be applied to the total number of
galaxies $N$ equal to 0.6. This means that our statistical
approximation carried an uncertainty that is equivalent to the
uncertainty of having $\sim\,60\%$ of the galaxies with full
information.

%%%%%%%%%%%%%%%%%%%%%%%%%%%%%%
% Table simulations
%%%%%%%%%%%%%%%%%%%%%%%%%%%%%%

\begin{table*}[!h]
  \centering
  \begin{tabular}{ccccccccc}

  Sample & $N$ &  $p_{input}$ & 
\multicolumn{2}{c}{$p_+$} & 
\multicolumn{2}{c}{$p_-$} & 
\multicolumn{2}{c}{$p_{output}$} \\
\cline{4-9}
\multicolumn{3}{c}{} &
$\langle p_+ \rangle$ &
$\sigma_{p_+}$ &
$\langle p_- \rangle$ &
$\sigma_{p_-}$ &
$\langle p_{output} \rangle$ &
$\sigma_{p_{output}}$ \\
\hline\hline
A &   11060 & 0.50 &  0.664 &  0.008 &  0.665 &  0.008 &  0.437 &  0.010 \\
 A &   11060 & 0.75 &  0.841 &  0.010 &  0.842 &  0.009 &  0.746 &  0.012 \\
 A &   11060 & 0.90 &  0.938 &  0.010 &  0.938 &  0.010 &  0.900 &  0.013 \\
 A &   11060 & 1.00 &  1.000 &  0.011 &  1.000 &  0.011 &  1.000 &  0.014 \\
 A &   11060 & 1.10 &  1.060 &  0.012 &  1.060 &  0.012 &  1.102 &  0.016 \\
 A &   11060 & 1.25 &  1.143 &  0.012 &  1.142 &  0.012 &  1.256 &  0.018 \\
 A &   11060 & 1.50 &  1.272 &  0.013 &  1.272 &  0.014 &  1.542 &  0.025 \\
\hline
 B &     179 & 0.50 &  0.642 &  0.060 &  0.644 &  0.059 &  0.486 &  0.059 \\
 B &     179 & 0.75 &  0.833 &  0.071 &  0.830 &  0.075 &  0.753 &  0.082 \\
 B &     179 & 0.90 &  0.938 &  0.081 &  0.938 &  0.085 &  0.907 &  0.101 \\
 B &     179 & 1.00 &  0.999 &  0.088 &  1.004 &  0.086 &  1.004 &  0.107 \\
 B &     179 & 1.10 &  1.073 &  0.091 &  1.072 &  0.091 &  1.115 &  0.121 \\
 B &     179 & 1.25 &  1.162 &  0.101 &  1.155 &  0.100 &  1.258 &  0.140 \\
 B &     179 & 1.50 &  1.305 &  0.113 &  1.313 &  0.110 &  1.573 &  0.561 \\
\hline
  \end{tabular}
  \caption{Results of several simulations to test the validity and
    robustness of our statistical correction. Two samples with different
    number of galaxies are shown: Sample A is made of
    galaxies in shells of $4\,\hmpc$ around voids with $R>10\,\hmpc$
    and Sample B is made of galaxies in shells of $3\,\hmpc$ around
    voids with $R>16\,\hmpc$. Each row corresponds to a set of 1000
    realizations in which to each real 
    galaxy a synthetic spin vector was assigned following the
    theoretical distribution given by Equation~(\ref{eq:pcostheta}) with
    a $p=p_{input}$. $p_+$ and $p_-$ 
    are the values of $p$ obtained when fixing the sign of
    $\zeta$. $p_{output}$ is the final value 
    after applying the statistical correction. For each parameter
    ($p_+$, $p_-$, $p_{output}$),
    the mean and the standard deviation of the 1000 realizations are
    shown. See text for more details. 
    \label{tab:p_check}
  }
\end{table*}

%%%%%%%%%%%%%%%%%%%%%%%%%%%%%%%%%%%%%%%%%%%%%%%%%%%%%%%%%%%%%%%%%%%%%%

\section{Computation of the uncertainties in
  $\meanabscostheta$ and $p$\label{app:uncertainties_p}}

In this section we present the expressions used to compute the
uncertainties in $\meanabscostheta$ and $p$, in the general
case.

The standard deviation of $\meanabscostheta$, \sigmameanabscostheta,
is computed as $\sigmacostheta/\sqrt{N_g}$, where $N_g$ is the total
number of galaxies.  $\sigmacostheta$ is the root mean square of
$\abscostheta$ for the distribution given by
Equation~\ref{eq:pcostheta}. Computing $\sigmacostheta$ analytically,
we find the following expressions for \sigmameanabscostheta\ depending
on the value of $p=\meanabscostheta^{-1}-1$, :
\begin{eqnarray}
  \sigmameanabscostheta & = &
  \frac{1}{\sqrt{N_g}}\sqrt{\frac{p}{(p^2-1)^{3/2}}\ln(p+\sqrt{p^2-1})-\frac{1}{p^2-1}-\frac{1}{1+p^2}};
  \quad p>1   \label{eq:err_cos_theta_1}\\
  \sigmameanabscostheta & = &
  \frac{1}{\sqrt{N_g}}\frac{1}{\sqrt{12}};
  \quad p=1 \label{eq:err_cos_theta_2}\\
  \sigmameanabscostheta & = &
  \frac{1}{\sqrt{N_g}}\sqrt{\frac{1}{1-p^2}-\frac{p}{(1-p^2)^{3/2}}\arcsin(\sqrt{1-p^2})-\frac{1}{1+p^2}};
  \quad p<1 \label{eq:err_cos_theta_3}
\end{eqnarray}

When using the method described in Appendix~\ref{app:robustness}, we
have some uncertainty in the direction of the spin, but we find that
the errors are well described by the above expressions using 0.6 times
the number of galaxies in the place of $N_g$ (see
Table~\ref{tab:p_check}).

Since the distribution of $p$ is not Gaussian, we can compute the
value $p$ and the limits of the $1\sigma$ confidence interval ($p_{-\sigma}$,
$p_{+\sigma}$) with the next expressions:
\begin{eqnarray}
  p & = & \frac{1}{\meanabscostheta}-1\\
p_{-\sigma}& = &
\frac{1}{\meanabscostheta+\sigma_{\meanabscostheta}}
  -1 \label{eq:p_errors_1}\\
p_{+\sigma}& = &
\frac{1}{\meanabscostheta-\sigma_{\meanabscostheta}}
  -1 \label{eq:p_errors_2}
\end{eqnarray}

\acknowledgements

This work has been supported by the Programa Nacional de Astronom\'{\i}a y
Astrof\'{\i}sica of the Spanish Ministry of Science and Innovation under
grant AYA2010-21322-C03-02.  J.V. acknowledges a post-doc fellowship
from the Spanish Ministry of Science and Innovation under the
programs 3I2005 and 3I2406.

Funding for the SDSS and SDSS-II has been provided by the Alfred
P. Sloan Foundation, the Participating Institutions, the National
Science Foundation, the U.S. Department of Energy, the National
Aeronautics and Space Administration, the Japanese Monbukagakusho, the
Max Planck Society, and the Higher Education Funding Council for
England. The SDSS Web Site is http://www.sdss.org/.

The SDSS is managed by the Astrophysical Research Consortium for the
Participating Institutions. The Participating Institutions are the
American Museum of Natural History, Astrophysical Institute Potsdam,
University of Basel, University of Cambridge, Case Western Reserve
University, University of Chicago, Drexel University, Fermilab, the
Institute for Advanced Study, the Japan Participation Group, Johns
Hopkins University, the Joint Institute for Nuclear Astrophysics, the
Kavli Institute for Particle Astrophysics and Cosmology, the Korean
Scientist Group, the Chinese Academy of Sciences (LAMOST), Los Alamos
National Laboratory, the Max-Planck-Institute for Astronomy (MPIA),
the Max-Planck-Institute for Astrophysics (MPA), New Mexico State
University, Ohio State University, University of Pittsburgh,
University of Portsmouth, Princeton University, the United States
Naval Observatory, and the University of Washington.
\end{document}